# On a formula of spin sums, Eisenstein-Kronecker series in higher genus Riemann surfaces

A.G. Tsuchiya

Abstract

We discuss a decomposition formula of simple products of fermion correlation functions with cyclic constrains and its applications to spin sums of super string amplitudes. Based on some facts which are noted or derived in this paper, we propose a candidate of the form of this decomposition formula for some of higher genus cases which includes genus 2 case. Although we had to use several conjectures and assumptions due to unsolved mathematical difficulties, the method described in the text may be an efficient way to obtain the decomposition formula in some of higher genus cases. In particular, for those cases, we propose a concrete method to sum over non-singular even spin structures for the product of arbitrary number of the fermion correlation functions with cyclic constraints in super string amplitudes.

We also propose an explicit generalization of Eisenstein-Kronecker series to the higher genus cases in the process of considerations above.

## 1. Genus one

### Introduction

What we consider in this paper is a simple product of fermion correlation functions

$S_\delta(z_1, z_2)S_\delta(z_2, z_3)\ldots\ldots S_\delta(z_N, z_{N+1})$ with a constraint $z_{N+1} = z_1$.

Here $S_\delta(z, w) = \frac{\theta[\delta](A(z-w))}{\theta[\delta](0)E(z,w)}$ is the fermion correlation function, $E(z, w) = \frac{\theta[\nu](A(z-w))}{h_\nu(z)h_\nu(w)}$ is the Prime form of genus g Riemann surfaces, and $A(z-w) \equiv (\int_w^z \omega_i) \in C^g$ represents Abel map for $i = 1,2, \ldots g$. $\delta$ represents non-singular even spin structure of the fermion correlation function.

The variables $z_1, z_2, \ldots\ldots z_N \in C^1$ are the inserting points of external particles (bosons) on Riemann surfaces. The constraint $z_{N+1} = z_1$ sometimes naturally occurs in calculating contractions of correlation functions using Wick theorem.

In this chapter we review the case of genus 1 as simply as possible, in a form that may be useful for generalising to cases of higher genus. Many of the details will be analogously repeated in the following chapters.

At genus one, we discuss $\Pi_{i=1}^N S_\delta(z_i - z_{i+1})$ under the constraint $z_{N+1} = z_1$ where

$$S_\delta(z - w) = \frac{{\theta_1}^{(1)}(0)\theta_{\delta+1}(z-w)}{\theta_{\delta+1}(0)\theta_1(z-w)} \tag{1.1}$$

for $\delta = 1,2,3$ . The notations are summarised in Appendix A.

We define $x_1 \equiv z_1 - z_2$, $x_2 \equiv z_3 - z_2$. ...., $x_N \equiv z_N - z_1$ , and hence

$$\sum_{i=1}^N \quad x_i = 0. \tag{1.2}$$

The type of the condition (1.2) appears naturally when loop correlation functions between any number of Kac Moody currents are considered. Therefore, when applied to superstring amplitudes, in heterotic string models we have to deal with the case of (1.2) when considering spin sums if the RNS formalism is adopted.

String amplitudes are sums of combinations of the products of fermion field Wick contractions and boson field contractions in the vertex operators. As for the fermion field contractions, for each Wick contraction, appropriate partition functions $Z_\delta$ , which are functions only of the moduli parameter tau, have to be multiplied and summed over the three different boundaries

$$\sum_{\delta=1,2,3} Z_\delta \prod_{i=1}^N S_\delta(x_i) \quad . \tag{1.3}$$

In these spin sum calculations, the boson field contractions can all be skipped.

Eq. (1.3) is the building block as a contribution from the Wick contraction of the fermion correlation functions in the super string amplitudes.
In the heterotic strings, the partition functions $Z_\delta$ take the form, up to the spin structure independent factors,

$$Z_\delta^{H,m} = [\theta_{\delta+1}(0)]^{8m} \qquad m = 1,2 \tag{1.4}$$

Another important type of example is the case of calculating the parity conserving part of the super string amplitudes of external massless bosons, with $N$ number of zero-pictured vertex operators which have the form

$$V(\zeta,k) = (\,\zeta\cdot\dot{X} - \zeta\cdot\psi\;\,k\cdot\,\psi\,)\,e^{k\cdot X} \tag{1.5}$$

where $\zeta, k$ are the polarization vector and momentum respectively.
In this calculation, the non-zero contributions of the Wick contractions of the fermion correlation functions are naturally restricted to only the case with condition (1.2) for arbitrary $N$, for the genus one case.
In this case, for the type I and type II superstrings, the partition function $Z_\delta$ is uniquely determined after incorporating the ghost and superghost mode contributions as

$$Z_\delta^B = (-1)^\delta\left[\frac{\theta_{\delta+1}(0)}{\theta_1^{(1)}(0)}\right]^4 \tag{1.6}$$

At the level of genus one, there is an efficient decomposition formula to perform the spin sum for arbitrary $N$.
For this purpose, we introduce a function $V$ defined as [1]

$$V_p(x_1,x_2,\ldots\ldots,x_N) \stackrel{\text{def}}{=} \frac{1}{p!}\frac{\partial^p}{\partial\alpha^p}exp(\sum_{i=1}^{N} ln\,\vartheta_1\,(x_i+\alpha) - \sum_{i=1}^{N} ln\,\vartheta_1\,(x_i) + N\sum_{k=1}^{\infty}\frac{\alpha^{2k}}{2k}G_{2k}(\tau))\bigg|_{\alpha=0} \tag{1.7}$$

where $\alpha$ is a common dummy parameter. Since only one parameter $\alpha$ relates to each of $x_i$ in linear form $x_i+\alpha$ in each function, after any number of differentiations with respect to $\alpha$ and setting it equal to zero, the differentiations become equivalent to those with $x_i$ s in its final form. $G_{2k}(\tau)$ is holomorphic Eisenstein series, defined from $k=1$ as follows :

$$G_{2k}(\tau) \equiv \textstyle\sum_{(m,n)\neq(0,0)}(n\tau+m)^{-2k}, \qquad (m,n) \in Z^2 \quad . \tag{1.8}$$

---

[1] This function $V$ is introduced in the first version of [2] and later precisely corrected in [3]. The form of $V$ in (1.7) is equivalent to that in [3] including all numerical factors.

We regard the case $k = 1$, $G_2(\tau)$ , is equal to

$$G_2(\tau) = 2\pi^2 B_2 - 8\pi^2 \sum_{n=1}^{\infty} \frac{nq^n}{1-q^n} \tag{1.9}$$

where $B_2$ is a Bernoulli number $\frac{1}{6}$ , $q = \exp(2\pi i\tau)$ . See Appendix A for notations.

We may write $V_p(x_1, x_2, \ldots\ldots, x_N)$ as $V_p$ if and only if the number of variables is $N$, and also write

$$P^{(n)}(\alpha) \stackrel{\text{def}}{=} \frac{\partial^n}{\partial\alpha^n} P(\alpha),\ P^{(-2)}(\alpha) \stackrel{\text{def}}{=} 1,\ P^{(-1)}(\alpha) \stackrel{\text{def}}{=} 0,\ P^{(0)}(\alpha) \stackrel{\text{def}}{=} P(\alpha),\ V_0(x_1, x_2, \ldots\ldots, x_N) \stackrel{\text{def}}{=} 1 \tag{1.10}$$

where $P$ is the Weierstrass Pe function.

Under the condition $\sum_{i=1}^{N} x_i = 0$, the product $\prod_{i=1}^{N} S_\delta(x_i)$ can be decomposed as

$$\prod_{i=1}^{N} S_\delta(x_i) = \sum_{K=0}^{\left[\frac{N}{2}\right]} H_{N,N-2K}\, P^{(2K-2)}(\omega_\delta)$$

$$( = H_{N,N} + H_{N,N-2}P(\omega_\delta) + H_{N,N-4}P^{(2)}(\omega_\delta) + \cdots + H_{N,N-2\left[\frac{N}{2}\right]} P^{\left(2\left[\frac{N}{2}\right]-2\right)}(\omega_\delta) ) \tag{1.11}$$

where

$$H_{N,N-2K} = \frac{1}{(2K-1)!} V_{N-2K} \qquad (\text{for } \mathrm{K} > 0.) \tag{1.12}$$

For $K = 0$, $H_{N,N}$ has different form as

$$H_{N,N} = V_N - \sum_{K=2}^{\left[\frac{N}{2}\right]} G_{2K} \cdot V_{N-2K} \quad . \tag{1.13}$$

The functions $H_{N,M}$ were first introduced in [5]. Here the $H_{N,M}$ are defined as the expansion coefficients in (1.24) later, and these forms are given as functions of $V$.

The variables $\omega_\delta$ are the half periods at the genus 1 :

$$\omega_1 = \frac{1}{2}, \quad \omega_2 = -\frac{1+\tau}{2}, \quad \omega_3 = \frac{\tau}{2} \quad . \tag{1.14}$$

The main point of this formula is that the boundary dependence ( $\delta$ dependence) of the product of the fermion correlation functions is factored out as $P^{(2K-2)}(\omega_\delta)$ which does not depend on the vertex insertion points $x_1, x_2, \ldots x_N$ but only on the moduli parameter $\tau$ . Therefore, the whole of spin sum $\sum_{\delta=1,2,3} Z_\delta \prod_{i=1}^{N} S_\delta(x_i)$ can be simplified as

$$\sum_{\delta=1,2,3} Z_\delta\ P^{(2K-2)}(\omega_\delta) \tag{1.15}$$

for each of $K$ .

Even-order differentiations of the Pe function, $P^{(2K-2)}(\alpha)$ , can be represented by polynomials of the Pe function itself as in (A.16), and so $P^{(2K-2)}(\omega_\delta)$ can be written by branch points $e_\delta$ via $P(\omega_\delta) = e_\delta$ . On the other hand, the partition function $Z_\delta$ can also be expressed in terms of branch points. For example, $Z_\delta^B$ , $(-1)^\delta \left[\frac{\theta_{\delta+1}(0)}{\theta_1^{(1)}(0)}\right]^4$ can be written as

$$\frac{(e_2-e_3)}{(e_1-e_2)(e_2-e_3)(e_3-e_1)}, \frac{(e_3-e_1)}{(e_1-e_2)(e_2-e_3)(e_3-e_1)}, \frac{(e_1-e_2)}{(e_1-e_2)(e_2-e_3)(e_3-e_1)} \quad \text{for } \delta = 1,2,3 \text{ respectively.} \tag{1.16}$$

For an arbitrary value of $N$, multiplying the simple product of the fermion correlation functions by the partition functions $Z_\delta$ and performing a spin sum, it becomes a simple algebraic sum of the branch points, depending only on the moduli. For example, $\sum_{\delta=1,2,3} Z_\delta^B \ P^{(2K-2)}(\omega_\delta)$ has the form

$$\frac{1}{(2K-1)!}\frac{(e_2-e_3)P^{(2K-2)}(\omega_1)+(e_3-e_1)P^{(2K-2)}(\omega_2)+(e_1-e_2)P^{(2K-2)}(\omega_3)}{(e_1-e_2)(e_2-e_3)(e_3-e_1)} \quad . \tag{1.17}$$

The numerator contains the denominator for arbitrary values of $K$ , and the whole of (1.17) can always be written as symmetric functions of $e_1, e_2, e_3$ and hence by Eisenstein series [2][4].

**Proof of (1.11)**

We use a classical formula:

$$S_\delta(x) = exp(-\eta_\delta x)\frac{\sigma(x+\omega_\delta)}{\sigma(x)\sigma(\omega_\delta)} = exp(-\eta_\delta x)F(x,\omega_\delta,\tau) \tag{1.18}$$

where $F(z,\alpha,\tau)$ is the Eisenstein-Kronecker series introduced in [3] and defined as follows :

$$F(x,\alpha,\tau) \stackrel{\text{def}}{=} \frac{\theta_1^{(1)}(0)\theta_1(x+\alpha,\tau)}{\theta_1(x,\tau)\theta_1(\alpha,\tau)} \quad , \tag{1.19}$$

or use

$$\frac{\theta_1(x+\omega_\delta)}{\theta_1(\omega_\delta)} = exp(-\pi i \ a_\delta \ x) \ \frac{\theta_{\delta+1}(x)}{\theta_{\delta+1}(0)} \tag{1.20}$$

where $a_\delta$ are the following constants:

$$a_1 = 0, \quad a_2 = -1, \quad a_3 = 1 \quad . \tag{1.21}$$

Then, if we consider the product

$$\Pi_{i=1}^N \ F(x_i,\alpha,\tau) = \Pi_{i=1}^N \frac{\theta_1^{(1)}(0)\theta_1(x_i+\alpha,\tau)}{\theta_1(x_i,\tau)\theta_1(\alpha,\tau)} \quad , \tag{1.22}$$

we can say that

$$\Pi_{i=1}^N S_\delta(x_i) = \Pi_{i=1}^N \ F(x_i,\omega_\delta,\tau) = \Pi_{i=1}^N \frac{\theta_1^{(1)}(0)\theta_1(x_i+\omega_\delta,\tau)}{\theta_1(x_i,\tau)\theta_1(\omega_\delta,\tau)} \tag{1.23}$$

under the condition (1.2), since the exp factor in (1.20) becomes 1.

The product $\Pi_{i=1}^{N} F(x_i,\alpha,\tau)$ can be regarded as an elliptic function of the parameter of $\alpha$ because of the condition $\sum_{i=1}^{N} x_i = 0$, and so can be expanded by the derivatives of Weierstrass Pe function as

$$\Pi_{i=1}^{N} F(x_i,\alpha,\tau) = \sum_{M=0}^{N} H_{N,M}(x_i)P^{(N-2-M)}(\alpha) = H_{N,N}(x_i) + H_{N,N-2}(x_i)P(\alpha) + H_{N,N-3}(x_i)P^{(1)}(\alpha) + H_{N,N-4}(x_i)P^{(2)}(\alpha) + \cdots + H_{N,0}P^{(N-2)}(\alpha) \quad (1.24)$$

where $H_{N,M}(x_i)$ are expansion coefficients defined here which do not depend on $\alpha$. There is no term $H_{N,N-1}(x_i)$ because there should be no single order pole of $\alpha$ in an elliptic function of $\alpha$. The order of the pole in the last term $P^{(N-2)}(\alpha)$ is $N$, which is the same order of the poles on the right-hand side of (1.24) as a function of $\alpha$. There is no need to consider terms with higher order poles of $\alpha$.

Here we modify the logarithm of $\Pi_{i=1}^{N} F(x_i,\alpha,\tau)$. Using

$$\ln\theta_1(\alpha) = \ln\sigma(\alpha) - \frac{G_2}{2}\alpha^2 + \ln\theta_1^{(1)}(0) = \ln\alpha - \sum_{k=1}^{\infty}\frac{G_{2k}}{2k}\alpha^{2k} + \ln\theta_1^{(1)}(0) \quad (1.25)$$

( See Appendix A ),

the left-hand side of (1.24) can be written as

$$\Pi_{i=1}^{N} F(x_i,\alpha,\tau) = \frac{1}{\alpha^N} exp\left( \sum_{i=1}^{N} ln\,\vartheta_1(x_i+\alpha) - \sum_{i=1}^{N} ln\,\vartheta_1(x_i) + N\sum_{k=1}^{\infty}\frac{\alpha^{2k}}{2k}G_{2k}(\tau)\right), \quad (1.26)$$

where the factor $N\,ln\alpha$ in $ln\Pi_{i=1}^{N} F(x_i,\alpha,\tau)$ has been extracted from the exp function. The whole is already the same as the content of the $V_p$ function except the pole factor $\frac{1}{\alpha^N}$.

We multiply $\alpha^N$ on both sides of (1.24) and consider differentiating $M$ times and setting $\alpha = 0$. Since the Pe function has the following single term singularities:

$$P^{(n)}(\alpha) = \frac{(-1)^n(n+1)!}{\alpha^{n+2}} + (n+1)!\,G_{n+2}(\tau) + O(\alpha)\ldots\ldots, \quad (1.27)$$

The term $H_{N,M}(x_i)P^{(N-2-M)}(\alpha)$ in eq. (1.24) multiplied by $\alpha^N$ has the form

$$H_{N,M}(x_i)\{\,(-1)^{N-M}(N-M-1)!\,\alpha^M + (N-M-1)!\,G_{N-M}(\tau)\alpha^N + O(\alpha^{N+1})\} \quad (1.28)$$

On both sides of eq. (1.24) multiplied by $\alpha^N$, there is no singular term, and for a fixed value of $M$ ( first assume $M < N$) there is only one term of order $\alpha^M$ in the whole of the right-hand side of the equation, which is $H_{N,M}(x_i)\,(-1)^{N-M}(N-M-1)!\,\alpha^M$. Therefore, for the case $M < N$, differentiating the both sides of $M$ times and setting $\alpha = 0$ we have $H_{N,M}(x_i)$ :

$$H_{N,M}(x_i) = \frac{(-1)^{N-M-2}}{M!(N-M-1)!}\frac{\partial^M}{\partial\alpha^M}[\alpha^N\Pi_{i=1}^{N} F(x_i,\alpha,\tau)]\ \Big|_{\alpha=0} = \frac{(-1)^{N-M-2}}{(N-M-1)!}V_M(x_1,x_2,\ldots\ldots,x_N) \quad (1.29)$$

To obtain $H_{N,N}(x_i)$ ( $M = N$), we note that every term on the right-hand side of (1.24) after multiplying $\alpha^N$ has a constant term except for the case of $M = N-2$

$$H_{N,M}(x_i)\,(N-M-1)!\,G_{N-M}(\tau) \quad (1.30)$$

( $G_{N-M}(\tau) = 0$ if $N-M$ is odd.)

Differentiating the both sides of (1.24) $N$ times after multiplying by $\alpha^N$ , we have

$$H_{N,N}(x_i) = V_N(x_1,x_2,\ldots\ldots,x_N) - \sum_{M=0}^{N-2} H_{N,M}(x_i)\,(N-M-1)!\,G_{N-M}(\tau) \qquad (1.31)$$

Then, $\prod_{i=1}^{N} S_\delta(x_i)$ under the condition $\sum_{i=1}^{N} x_i = 0$ can be obtained by setting $\alpha$ equals to half period $\omega_\delta$ :

$$\prod_{i=1}^{N} S_\delta(x_i) = \prod_{i=1}^{N} F(x_i,\omega_\delta,\tau) = \sum_{M=0}^{N} H_{N,M}(x_i) P^{(N-M-2)}(\omega_\delta)$$

$$= \sum_{M=0}^{N-1} \frac{(-1)^{N-M-2}}{(N-M-1)!} V_M(x_1,x_2,\ldots\ldots,x_N) P^{(N-M-2)}(\omega_\delta) + H_{N,N}(x_i) \qquad (1.32)$$

Since $P^{(ODD)}(\omega_\delta) = 0$ , by defining a new variable $K$ as $N - M = 2K$, this can be simplified as:

$$\prod_{i=1}^{N} S_\delta(x_i) = V_N(x_i) - \sum_{K=2}^{\left[\frac{N}{2}\right]} G_{2K} \cdot V_{N-2K}(x_i) + \sum_{K=1}^{\left[\frac{N}{2}\right]} \frac{1}{(2K-1)!} V_{N-2K}(x_i) P^{(2K-2)}(\omega_\delta) \qquad (1.33)$$

where $P^{(-2)}(\alpha) \overset{\text{def}}{=} 1$, $P^{(-1)}(\alpha) \overset{\text{def}}{=} 0$, $P^{(0)}(\alpha) \overset{\text{def}}{=} P(\alpha)$ as defined above.
This has the same form as (1.11), (1.12), (1.13).

The factor $\frac{1}{(2K-1)!}$ in the last term of eq. (1.33) comes from the numerical factor in the expansion of Pe in (1.27), $(-1)^n(n+1)!$ . This $(2K-1)!$ is equal to the numerical coefficient in the highest degree term of the polynomial of the Pe function when $P^{(2K-2)}(\omega_\delta)$ is written by the Pe function.

In the calculations of external massless boson amplitudes in the Type I and Type II superstrings, $H_{N,N}$ does not contribute to the final form of the amplitudes after summing over spin structures for any $N$ , because $\sum_{\delta=1,2,3} Z_\delta^B = 0$ , and only $H_{N,N-2K}$ can have potential contributions for $K > 0$.
On the other hand, for heterotic string amplitudes for example, $H_{N,N}$ can contribute to the final form of the amplitudes because $\sum_{\delta=1,2,3} Z_\delta^{Hm}$ is non-zero. In this case, the exact form of $H_{N,N}$ (1.13) must be used to calculate $N$ point amplitudes at loop 1 level.
The additional terms of eq. (1.13), $\sum_{K=2}^{\left[\frac{N}{2}\right]} G_{2K} \cdot V_{N-2K}$ , have non-zero contributions even at the heterotic 4-point amplitudes, as will be seen in (1.58).

**The constant terms in the expansion of derivatives of odd theta function in genus one**

The spin structure sum method in the form (1.11) was discussed in [1] and [2], where $H_{N,N-2K}$ ( $K > 0$) and $H_{N,N}$ were obtained as the ratio of two determinants of elliptic functions, as in (A.22), (A23). It was desirable to rewrite these ratios of two determinants in more standard forms as polynomials of $\frac{\partial^k}{\partial x_i^k} ln\, \theta_1(x_i, \tau)$ . In [1][2], such a rewritten form of $H_{N,N-2K}$ ( $K > 0$) was described by pole structure considerations only up to $N = 7$, where the form of $V_p(x_1, x_2, \ldots\ldots, x_N)$ was inaccurate for $N > 7$. The authors of [3] wrote down a correct formula of $V_p(x_1, x_2, \ldots\ldots, x_N)$ for arbitrary $N$, in a form equivalent to (1.7). In their correct formula of $V_p$, although the product of fermion correlation functions with cyclic conditions $\prod_{i=1}^{N} S_\delta(x_i)$ has a pole of the form $\frac{1}{x_1 x_2 \cdots x_N}$, there are less singular pole terms which originate from the factor $\theta_1(\alpha, \tau)$ in the denominator of Eisenstein-Kronecker series (1.23) or equivalently from the terms $N \sum_{k=1}^{\infty} \frac{\alpha^{2k}}{2k} G_{2k}(\tau)$ in the form of (1.7).

We describe some simple calculations and observations for later convenience in Chapters 2 and 3 in the below.

For a single factor of $F(x, \alpha, \tau)$ , we modify as

$$F(x, \alpha, \tau) = \frac{1}{\alpha} exp( \; ln\, \sigma(x + \alpha, \tau) - ln\, \sigma(x, \tau) - \alpha x G_2 + \sum_{k=2}^{\infty} \frac{\alpha^{2k}}{2k} G_{2k}(\tau))$$

$$= \frac{1}{\alpha} exp( \; ln\, \theta_1(x + \alpha, \tau) - ln\, \theta_1(x, \tau) + \sum_{k=1}^{\infty} \frac{\alpha^{2k}}{2k} G_{2k}(\tau))$$

$$= \frac{1}{\alpha} exp( \sum_{k \geq 1} \frac{\alpha^k}{k!} \frac{\partial^k}{\partial \alpha^k} ln\, \theta_1(x + \alpha, \tau) \Big|_{\alpha=0} + \sum_{k=1}^{\infty} \frac{\alpha^{2k}}{2k} G_{2k}(\tau))$$

$$= \frac{1}{\alpha} exp( \sum_{k \geq 1} \frac{\alpha^k}{k!} \frac{\partial^k}{\partial x^k} ln\, \theta_1(x, \tau) + \sum_{k=1}^{\infty} \frac{\alpha^{2k}}{2k} G_{2k}(\tau)) \tag{1.34}$$

We have used the $k = 1$ term in the exp function $\frac{\alpha^2}{2} G_2(\tau)$ to express the sigma function by $ln\, \theta_1(x, \tau)$. $G_2(\tau)$ refers to the value of the zeta function at one of the half periods as $\eta_1 = \frac{G_2}{2} = \zeta(\omega_1)$ . Hence we have

$$\alpha^N \prod_{i=1}^{N} F(x_i, \alpha, \tau) = \; exp( \sum_{k=1}^{\infty} \frac{\alpha^k}{k!} \sum_{i=1}^{N} \{ \frac{\partial^k}{\partial x_i^k} ln\, \theta_1(x_i, \tau) + (k-1)! \; G_k \}) \tag{1.35}$$

where $G_k = 0$ if $k$ is odd.

Note that $(k-1)! \; G_k$ exactly cancels the constant term ( $(x_i)^0$ term) of the function $\frac{\partial^k}{\partial x_i^k} ln\, \theta_1(x_i, \tau)$ when it is Laurant expanded. We temporally call this $(x_i)^0$ term $-(k-1)! \; G_k$ "zero mode" of the $\frac{\partial^k}{\partial x_i^k} ln\, \theta_1(x_i, \tau)$ , just for short, for convenience in later chapters.

We define the “zero mode subtracted” functions $D_i^k \, ln\, \theta_1\,(x_i)$ as

$$D_i^k \, ln\, \theta_1\,(x_i) \stackrel{\text{def}}{=} \frac{\partial^k}{\partial x_i^k} ln\, \theta_1\,(x_i,\tau) + (k-1)!\, G_k \tag{1.36}$$

$D_i^k \, ln\, \theta_1\,(x_i)$ is itself equal to $\frac{\partial^k}{\partial x_i^k} ln\, \theta_1\,(x_i,\tau)$ if $k$ is odd.

In the calculation of $V$, only the function $D_i^k \, ln\, \theta_1\,(x_i)$ appears. This can be simply understood by looking at the form of $V$ as follows.
See the inside of the exp function of the definition of $V_p(x_1, x_2, \ldots\ldots, x_N)$ :

$$\sum_{i=1}^N ln\, \vartheta_1\,(x_i+\alpha) - \sum_{i=1}^N ln\, \vartheta_1\,(x_i) + N\sum_{k=1}^\infty \frac{\alpha^{2k}}{2k} G_{2k}(\tau) \tag{1.37}$$

The three sets of terms of sum in the exp function of $V$ all come from the same function, $ln\theta_1$ . The final set of terms $N\sum_{k=1}^\infty \frac{\alpha^{2k}}{2k} G_{2k}(\tau)$ comes from $-N\, ln\theta_1(\alpha,\tau)$ excluding terms that are singular at $\alpha = 0$. After differentiating $V_p$ any number of times (say $k$ times) with respect to $\alpha$ and setting $\alpha = 0$ , these terms leave only the $(\alpha)^0$ terms. The first term in the exp function, $\sum_{i=1}^N ln\, \vartheta_1\,(x_i+\alpha)$ , is the same form of the function and gives $\frac{\partial^k}{\partial x_i^k} ln\, \theta_1\,(x_i,\tau)$ . Another set of terms $-\sum_{i=1}^N ln\, \vartheta_1\,(x_i)$ is just to make the inside of the exp function zero after the calculation. Then it is obvious that the $(x_i)^0$ term in the expansion form of $\frac{\partial^k}{\partial x_i^k} ln\, \theta_1\,(x_i,\tau)$ should always be the same as the left function, the $(\alpha)^0$ term with opposite sign.

The reason why we describe this simple observation in detail is as follows.

As was pointed out in [5], the factor $\theta_1(\alpha,\tau)$ in the denominator on the right-hand side of the product of Eisenstein-Kronecker series

$$\prod_{i=1}^N F(x_i,\alpha,\tau) = \prod_{i=1}^N \frac{\theta_1^{(1)}(0)\theta_1(x_i+\alpha,\tau)}{\theta_1(x_i,\tau)\theta_1(\alpha,\tau)}$$

is a key to the residue condition

$$Res_{x_N=0}\ H_{N,M}(x_1, x_2, \ldots . x_N) = H_{N-1,M-1}(x_1, x_2, \ldots . x_{N-1})\,. \tag{1.38}$$

As we have seen, since the product of the fermion correlation functions is related to $\prod_{i=1}^N F(x_i,\alpha,\tau)$ as

$$\prod_{i=1}^N S_\delta(x_i) = \prod_{i=1}^N F(x_i,\omega_\delta,\tau) = \prod_{i=1}^N \frac{\theta_1^{(1)}(0)\theta_1(x_i+\omega_\delta,\tau)}{\theta_1(x_i,\tau)\theta_1(\omega_\delta,\tau)}\ ,$$

in the decomposition formula of $\prod_{i=1}^N S_\delta(x_i)$ , the effects of such $\theta_1(\alpha,\tau)$ in the denominator are all included in a form as $D_i^k \, ln\, \theta_1\,(x_i)$ . As we will see later, in some cases of higher genus, a similar formula holds, and hence the zero mode subtractions may also hold in these cases. The description here is for the preparation of such higher genus cases.

The set of these “zero modes” is obviously the set of all coefficients of non-singular terms in the expansion of $ln\,\theta_1\,(x,\tau)$ . In genus 1, these are $-(k-1)!\,G_k$ , essentially holomorphic Eisenstein series. The first one is $-G_2$ , which is equal to $-2\eta_1$ , the constant used in the definition of the Pe function.

It may be interesting to note that the “zero mode subtraction” does not only occur in the calculation of $V$ functions, but also in another place. We have seen that $H_{N,N}$ has additional terms $\sum_{K=2}^{\left[\frac{N}{2}\right]} G_{2K}\cdot V_{N-2K}$ , compared to $H_{N,N-2K}$ $(K>0)$ , as in (1.12) and (1.13). These additional terms come from the existence of constant terms ( $\alpha$ -independent terms, zero modes ) in the base functions of the expansion $P^{(k)}(\alpha)$ , as seen in the proof of (1.11). This can be seen by rewriting the form of (1.11) as (2.52) using the notation (1.36).

**Examples**

When we expand the right-hand side of (1.35), the coefficient of $\alpha^M$ for arbitrary $M$ can be written down explicitly by considering the form of the weight $M$ Schur polynomial. For any $y_i$ and $\Lambda_i$ the following equation holds:

$$\exp(y_1\Lambda+y_2\Lambda^2+\cdots+y_{n-1}\Lambda^{n-1})=\sum_{M=0}^{n-1} p_M(y)\Lambda^M \tag{1.39}$$

where $p_M(y)$ is a polynomial of $y_i$ s given as

$$p_M(y)=\sum_{m_1+2m_2+3m_3+\cdots Mm_M=M}\frac{{y_1}^{m_1}{y_2}^{m_2}\cdots{y_M}^{m_M}}{m_1!m_2!\ldots m_M!} \tag{1.40}$$

and the summation is over all possible combinations of non-negative integers of $m_1,m_{2,}\ldots.\,m_M$ .

Using this formula, we have

$$V_M(x_1,x_2,\ldots\ldots,x_N)$$

$$=\sum_{m_1+2m_2+3m_3+\cdots Mm_M=M}\frac{\left(\sum_{i=1}^N D_i^1\,ln\,\theta_1\,(x_i)\right)^{m_1}\left(\sum_{i=1}^N D_i^2\,ln\,\theta_1\,(x_i)\right)^{m_2}\ldots\left(\sum_{i=1}^N D_i^M\,ln\,\theta_1\,(x_i)\right)^{m_M}}{(1!)^{m_1}(2!)^{m_2}\cdots\cdots(M!)^{m_M}\,m_1!\,m_2!\ldots m_M!} \tag{1.41}$$

Here are some examples to be used later:

1) N=2 calculate $S_\delta(x_1)S_\delta(x_2)$ where $x_1+x_2=0$ :

We calculate the right-hand side of (1.11):

$$\sum_{K=0}^{\left[\frac{N}{2}\right]} H_{N,N-2K}\,P^{(2K-2)}(\omega_\delta)=V_2(x_1,x_2)\ +V_0(x_1,x_2)P(\omega_\delta)=V_2(x_1,x_2)\ +e_\delta \tag{1.42}$$

$V_2(x_1, x_2)$ has the possible combination $(m_1, m_2) = (2{,}0),\ (0{,}1)$ in (1.41), which gives

$$V_2(x_1, x_2) = \frac{1}{2}(D^2_{x_1}\, ln\,\theta_1\,(x_1) + D^2_{x_2}\, ln\,\theta_1\,(x_2)) + \frac{1}{2}\left(\partial^1_{x_1}\, ln\,\theta_1\,(x_1) + \partial^1_{x_2}\, ln\,\theta_1\,(x_2)\right)^2 \qquad (1.43)$$

Since

$$D^2_i\, ln\,\theta_1\,(x) \stackrel{\text{def}}{=} \frac{\partial^2}{\partial x^2} ln\,\theta_1\,(x, \tau) + G_2 = \ -P(x) \qquad (1.44)$$

and the term $\frac{1}{2}\left(\partial^1_{x_1}\, ln\,\theta_1\,(x_1) + \partial^1_{x_2}\, ln\,\theta_1\,(x_2)\right)^2$ vanishes after setting $x_1 + x_2 = 0$, we have, writing $x = x_1 = -x_2$ ,

$$S_\delta(x_1)S_\delta(x_2) = -\,(S_\delta(x))^2 = -\,P(x) + e_\delta \qquad (1.45)$$

which is a famous classical formula.

If we use Weierstrass relations

$$2\eta_1 = G_2 = -e_\delta - \frac{\theta^{(2)}_{\delta+1}(0)}{\theta_{\delta+1}(0)} \qquad (1.46)$$

(for any of $\delta = 1{,}2{,}3$ ) , eq. (1.45) is written as

$$(S_\delta(x))^2 = \ -\frac{\partial^2}{\partial x_i^2} ln\,\theta_1\,(x_i, \tau) + \frac{\theta^{(2)}_{\delta+1}(0)}{\theta_{\delta+1}(0)} \quad . \qquad (1.47)$$

This means that, since $e_\delta = -2\eta_1 - \frac{\theta^{(2)}_{\delta+1}(0)}{\theta_{\delta+1}(0)}$ , the spin structure dependence of the product $\Pi^N_{i=1} S_\delta(x_i)$ under the condition (1.2) is satisfied by only one kind of theta constants, $\frac{\theta^{(2)}_{\delta+1}(0)}{\theta_{\delta+1}(0)}$. There exists a natural generalisation of this formula in hyperelliptic cases of arbitrary genus cases when $\delta$ represents non-singular even characteristics.

2) N=3 $\quad S_\delta(x_1)S_\delta(x_2)S_\delta(x_3)$ where $x_1 + x_2 + x_3 = 0$

This product becomes

$$\sum_{K=0}^{\left[\frac{N}{2}\right]} H_{N,N-2K}\, P^{(2K-2)}(\omega_\delta) = V_3(x_1, x_2, x_3) \ + V_1(x_1, x_2, x_3) P(\omega_\delta) \qquad (1.48)$$

$V_3(x_1, x_2, x_3)$ has possible combinations $(m_1, m_2, m_3) = (3{,}0{,}0),\ (1{,}1{,}0), (0{,}0{,}1)$ .

Then

$$V_3(x_1, x_2, x_3)$$

$$= \frac{1}{6}(\textstyle\sum^3_{i=1} \partial^3_{x_i}\, ln\,\theta_1\,(x_i)) + \frac{1}{2}\left[\ \sum^3_{i=1} \partial^1_{x_i} ln\,\theta_1\,(x_i)\ \right]\left[\ \sum^3_{i=1}\ D^2_{x_i}\ ln\,\theta_1\ \ \right] \ + \frac{1}{6}(\textstyle\sum^3_{i=1} \partial^1_{x_i}\, ln\,\theta_1\,(x_i))^3 \qquad (1.49)$$

and

$S_\delta(x_1)S_\delta(x_2)S_\delta(x_3)$

$$= \frac{1}{6}\left(\sum_{i=1}^{3} \partial_{x_i}^3 \ln\theta_1(x_i)\right) + \frac{1}{2}\left[\sum_{i=1}^{3} \partial_{x_i}^1 \ln\theta_1(x_i)\right]\left[\sum_{i=1}^{3} D_{x_i}^2 \ln\theta_1\right]$$

$$+\frac{1}{6}\left(\sum_{i=1}^{3} \partial_{x_i}^1 \ln\theta_1(x_i)\right)^3 + \left[\sum_{i=1}^{3} \partial_{x_i}^1 \ln\theta_1(x_i)\right] e_\delta \qquad (1.50)$$

For N=3, under the condition $x_1 + x_2 + x_3 = 0$, the following identity also holds：

$$\left[\sum_{i=1}^{3} \partial_{x_i}^1 \ln\theta_1(x_i)\right]^2 = -\sum_{i=1}^{3} D_{x_i}^2 \ln\theta_1(x_i) = \sum_{i=1}^{3} P(x_i) \qquad (1.51)$$

We can rewrite the $V_3(x_1, x_2, x_3)$ as

$$V_3(x_1, x_2, x_3) = \frac{1}{6}\left(\sum_{i=1}^{3} \partial_{x_i}^3 \ln\theta_1(x_i)\right) + \frac{1}{3}\left[\sum_{i=1}^{3} \partial_{x_i}^1 \ln\theta_1(x_i)\right]\left[\sum_{i=1}^{3} D_{x_i}^2 \ln\theta_1(x_i)\right]. \qquad (1.52)$$

Then eq. (1.50) can also be written as

$S_\delta(x_1)S_\delta(x_2)S_\delta(x_3)$

$$= \frac{1}{6}\left(\sum_{i=1}^{3} \partial_{x_i}^3 \ln\theta_1(x_i)\right) + \frac{1}{3}\left[\sum_{i=1}^{3} \partial_{x_i}^1 \ln\theta_1(x_i)\right]\left[\sum_{i=1}^{3} D_{x_i}^2 \ln\theta_1(x_i)\right]$$

$$+\left[\sum_{i=1}^{3} \partial_{x_i}^1 \ln\theta_1(x_i)\right] e_\delta \qquad (1.53)$$

3) N=4 $S_\delta(x_1)S_\delta(x_2)S_\delta(x_3)S_\delta(x_4)$ where $x_1 + x_2 + x_3 + x_4 = 0$

$$\sum_{K=0}^{\left[\frac{N}{2}\right]} H_{N,N-2K}\, P^{(2K-2)}(\omega_\delta) = V_4(x_1, x_2, x_3, x_4) - G_4 + V_2(x_1, x_2, x_3, x_4)P(\omega_\delta) + \frac{1}{3!}P^{(2)}(\omega_\delta) \qquad (1.54)$$

$V_4(x_1, x_2, \ldots. x_4)$ has possible combinations $(m_1, m_2, m_3, m_4) = (4,0,0,0)$, $(2,1,0,0), (0,0,2,0), (1,0,1,0), (0,0,0,1)$ .

Then

$V_4(x_1, x_2, \ldots. x_4)$

$$= \frac{1}{4!}\left[\sum_{i=1}^{4} \partial_{x_i}^1 \ln\theta_1(x_i)\right]^4 + \frac{1}{2!2!}\left[\sum_{i=1}^{4} \partial_{x_i}^1 \ln\theta_1(x_i)\right]^2 \left[\sum_{i=1}^{4} D_{x_i}^2 \ln\theta_1(x_i)\right]$$

$$+\frac{1}{3!}\left[\sum_{i=1}^{4} \partial_{x_i}^1 \ln\theta_1(x_i)\right]\left[\sum_{i=1}^{4} \partial_{x_i}^3 \ln\theta_1(x_i)\right] + \frac{1}{2!2!}\left[\sum_{i=1}^{4} D_{x_i}^2 \ln\theta_1(x_i)\right]^2$$

$$+\frac{1}{4!}\left[\sum_{i=1}^{4} D_{x_i}^4 \ln\theta_1(x_i)\right]. \qquad (1.55)$$

We also have

$$V_2(x_1, x_2, \ldots. x_4) = \frac{1}{2}\left(\sum_{i=1}^{4} D_{x_i}^2 \ln\theta_1(x_i)\right) + \frac{1}{2}\left(\sum_{i=1}^{4} \partial_{x_i}^1 \ln\theta_1(x_i)\right)^2. \qquad (1.56)$$

The last term of (1.55), $+\frac{1}{4!}\left[\sum_{i=1}^{4} D_{x_i}^4 \ln\theta_1(x_i)\right]$, includes $+G_4$ when considering $D_{x_i}^4$ .

In $H_{4,4}$ , this $+G_4$ is cancelled by the contribution $-\sum_{K=2}^{\left[\frac{N}{2}\right]} G_{2K} \cdot V_{N-2K}$ of the additional terms in (1.13) which is equal to $-G_4$ in eq. (1.13).

As can be seen from (1.12), for any $N > 4$

$$H_{N,4} = \frac{1}{(N-5)!} V_4(x_1, x_2, \ldots\ldots, x_N) \quad , \tag{1.57}$$

as in eq. (3.52) of [5], but only for $N = 4$, because of eq. (1.13),

$$H_{4,4} = V_4 - G_4 \ . \tag{1.58}$$

This $H_{4,4}$ is used in the calculations of the four-point heterotic massless boson amplitudes. In higher point heterotic amplitudes, general forms of $H_{N,N}$ (as well as $H_{N,N-2K}$ ) are sufficient to sum over spin structures for arbitrary $N$.

For N=4, under the condition $x_1 + x_2 + x_3 + x_4 = 0$ , the following identity also holds:

$$[\sum_{i=1}^{4} \partial_{x_i}^1 \ln\theta_1(x_i) \ ]^3 = -3\,[\sum_{i=1}^{4} \partial_{x_i}^1 \ln\theta_1(x_i) \ ][\sum_{i=1}^{4} D_{x_i}^2 \ln\theta_1(x_i) \ ]$$
$$-\sum_{i=1}^{4} \partial_{x_i}^3 \ln\theta_1(x_i) \tag{1.59}$$

Using this, we can rewrite $V_4(x_1, x_2, \ldots. x_N)$ as

$$V_4(x_1, x_2, \ldots. x_4) \ = +\frac{1}{8}\,[\sum_{i=1}^{4} \partial_{x_i}^1 \ln\theta_1(x_i)]^2 \ [\ \sum_{i=1}^{4} D_{x_i}^2 \ln\theta_1(x_i)]$$

$$+\frac{1}{8}\ [\sum_{i=1}^{4} \partial_{x_i}^1 \ln\theta_1(x_i)]\ \ [\ \sum_{i=1}^{4} \partial_{x_i}^3 \ln\theta_1(x_i)] \ + \frac{1}{8}\ [\sum_{i=1}^{4} D_{x_i}^2 \ln\theta_1(x_i)]^2$$
$$+\frac{1}{4!}[\sum_{i=1}^{4} D_{x_i}^4 \ln\theta_1(x_i)] \quad . \tag{1.60}$$

As far as the fermion field correlation functions are concerned, we have to set the common parameter $\alpha$ to $\omega_\delta$ at the end in the whole the product

$$\Pi_{i=1}^{N}\ F(x_i, \alpha, \tau) = \Pi_{i=1}^{N} \frac{\theta_1^{(1)}(0)\theta_1(x_i+\alpha,\tau)}{\theta_1(x_i,\tau)\theta_1(\alpha,\tau)} \quad ,$$

to derive a decomposition formula for the product of the fermion correlation functions. Setting $\alpha = 0$ is only to obtain the coefficients $H_{N,M}$ . In the decomposition formula (1.11), (1.12), (1.13), due to the simple pole structure of the Pe function in the base functions of the expansion, $V$ has the same form as that of the Taylor expansion of $\alpha^N \Pi_{i=1}^{N}\ F(x_i, \alpha, \tau)$ , each term of which is multiplied by the elliptic effects $\frac{1}{(2K-1)!} P^{(2K-2)}(\omega_\delta)$ . We note that the spin structure dependent parts come from the base functions of the expansion, $P^{(2K-2)}(\alpha)$ , by setting $\alpha = \omega_\delta$ .

## 2. higher genus

### 2.1 General considerations

#### Notations

Consider the branch points of the curve, $e_1, e_2, \ldots\ldots e_{2g+2}$, and fix the value of $e_{2g+2}$ at $\infty$. These branch points $e_1, e_2, \ldots\ldots e_{2g+2}$ are from left to right, and there are cuts between $e_1$ and $e_2$; $e_3$ and $e_4$ ; ... $e_{2g+1}$ and $e_{2g+2} = \infty$ . (Please see Figure 1 of [11].)

What we consider here is a simple product of fermion correlation functions

$$S_\delta(z_1, z_2) S_\delta(z_2, z_3) \ldots\ldots S_\delta(z_N, z_{N+1}) \text{ with the constraint } z_{N+1} = z_1. \qquad (1.2)$$

Here $S_\delta(z, w) = \frac{\theta[\delta](A(z-w))}{\theta[\delta](0) E(z,w)}$ is the Szego kernel, $E(z, w) = \frac{\theta[\nu](A(z-w))}{h_\nu(z) h_\nu(w)}$ is the prime form of genus g Riemann surfaces, and $A(z-w) \equiv (\int_w^z \omega_i) \in C^g$ represents the Abel map. $z_1, z_2, \ldots\ldots z_N \in C^1$ , and . $\omega_i$ are holomorphic 1-forms, $i = 1,2, \ldots g$.

The variables $z_1, z_2, \ldots\ldots z_N$ are the insertion points of external particles (bosons) on Riemann surfaces.

The Jacobi theta function in higher genus is defined in a standard notation:

$$\theta \begin{bmatrix} a \\ b \end{bmatrix} (u,\ \Omega) = \sum_{n \in Z^g} \quad exp\{ i\pi (n+a)^t T (n+a) + 2\pi i (n+a)(u+b) \} \qquad (2.1)$$

$$u \in C^g \qquad a, b \in \frac{1}{2} Z^g \quad . \qquad (2.2)$$

The theta function is called odd or even depending on whether the inner product $4a \cdot b$ is even or odd. The matrix $T$ is defined in (2.6) below.

Let $A_1, A_2, \ldots A_g$ and $B_1, B_2, \ldots. B_g$ be a canonical homology basis. We choose canonical holomorphic differentials of the first kind $\omega_1, \omega_2, \ldots \omega_g$ and associated meromorphic differentials of the second kind $r_1, r_2, \ldots r_g$ . The periods are given as

$$M_{IJ} = \frac{1}{2} \oint_{A_I} \omega_J \qquad \widehat{M_{IJ}} = \frac{1}{2} \oint_{B_I} \omega_J \qquad (2.3)$$

$$\eta_{IJ} = -\frac{1}{2} \oint_{A_I} r_J \qquad \hat{\eta}_{IJ} = -\frac{1}{2} \oint_{B_I} r_J \quad . \qquad (2.4)$$

In the following, $\eta_{IJ}$ , which corresponds to $\eta_1$ in the genus 1, will play an important role. The notations (2.3), (2.4) are the same as in ref. [13].

**Classification of spin structures**

As far as fermion correlation functions are concerned, in the case of $g = 2$ for example, the theta function which has the set of the following characteristics may play the same role as $\ln\theta_1(x)$ of the genus 1:

$$\theta\begin{bmatrix}\Delta^a\\ \Delta^b\end{bmatrix}(u) \quad \text{where} \quad \Delta^a = \left(0,\ \tfrac{1}{2}\right), \quad \Delta^b = \left(\tfrac{1}{2},\ \tfrac{1}{2}\right). \tag{2.5}$$

There is natural generalisation for any higher genus in the hyperelliptic case, but as far as I know, the Laurent expansion form of $\ln\theta\begin{bmatrix}\Delta^a\\ \Delta^b\end{bmatrix}(u)$ is not known in any genus except $g = 1$.

Consider the abelian image of each of the branch points as $\Omega_j \equiv (2M)^{-1}\int_{\infty}^{(e_j,0)}\omega_I \in C^g$. This is the basic tool for considering half periods in Riemann surfaces of higher genus. Each of $\Omega_j$ is the $g$ component vector for all of holomorphic one-forms $\omega_I$, $I = 1,2,\dots g$. The index $j$ is attached to each of the branch points $e_j$, $j = 1,2,\dots 2g+1$. For $j = 2g+2$, the integral is zero. The results of the integrals have the form

$$\Omega_j = (2M)^{-1}\int_{\infty}^{e_j}\omega_I = a_j\,T + b_j\,, \qquad T \stackrel{\text{def}}{=} M^{-1}\widehat{M} \tag{2.6}$$

Both of $a_j$ and $b_j$ for each $j$ are vectors of dimension $g$ with all elements either of 0 or $\frac{1}{2}$. $\Omega_j$ are also dimension $g$ vectors whose components depends on holomorphic 1 forms $\omega_I$ for $I = 1,2,\dots g$. Strictly speaking $\Omega$ should be written as $\Omega_j^I$, but we will sometimes abbreviate the index $I$.

Please note also that the result of the integral is not written as $b_j\ T + a_j$, but as $a_j\ T + b_j$.

Since $a_j$ and $b_j$ are often used as characteristics of theta functions, we call the pair even or odd when the inner product $4a\cdot b$ is even or odd.

For example, in the case of $g = 2$, since there are $2g+2 = 6$ branch points,

$$a_1 = \tfrac{1}{2}(0,\ 0)\ \ b_1 = \tfrac{1}{2}(1,\ 0);\quad a_2 = \tfrac{1}{2}(1,\ 0)\ \ b_2 = \tfrac{1}{2}(1,\ 0);\quad a_3 = \tfrac{1}{2}(1,\ 0)\ \ b_3 = \tfrac{1}{2}(0,\ 1)$$
$$a_4 = \tfrac{1}{2}(1,\ 1)\ \ b_4 = \tfrac{1}{2}(0,\ 1);\quad a_5 = \tfrac{1}{2}(1,\ 1)\ \ b_5 = \tfrac{1}{2}(0,\ 0);\quad a_6 = \tfrac{1}{2}(0,\ 0)\ \ b_6 = \tfrac{1}{2}(0,\ 0). \tag{2.7}$$

We are also interested in the following combinations of $\Omega_k$:

Choose any of combination of 2 indices $i, j$ from the non-zero $2g+1$ number of $\Omega_k$ in the case of genus $g$, and define

$$\Omega_{ij} = \Omega_i + \Omega_j = (a_i + a_j)\,T + (b_i + b_j). \tag{2.8}$$

Similarly, we define $\Omega_{ijk}$, which is any three combinations of $\Omega_j$, that is, $(a_i + a_j + a_k)T + (b_i + b_j + b_k)$.

Also, we define $\Omega_{ijkl}$ , $\Omega_{ijklm}$ , .... .

We define the following sum, choosing only even numbers of $k$ in $\Omega_k$ and summing all of them in genus $g$ :

$$R \overset{\text{def}}{=} \Omega_2 + \Omega_4 + \Omega_6 + \cdots \Omega_{2g} = \left(\sum_{i=1}^{g} a_{2i}\right)T + \left(\sum_{i=1}^{g} b_{2i}\right) . \qquad (2.9)$$

$R$ is called the vector of Riemann constants with the base point $\infty$ .

We define $\Delta^a \overset{\text{def}}{=} \sum_{i=1}^{g} a_{2i}$ , $\Delta^b \overset{\text{def}}{=} \sum_{i=1}^{g} b_{2i}$ .

In genus 2, $R$ becomes

$$R = \Omega_2 + \Omega_4 = (a_2 + a_4)T + (b_2 + b_4) = \Delta^a T + \Delta^b \quad \text{and so} \quad \Delta^a = \left(0, \tfrac{1}{2}\right), \Delta^b = \left(\tfrac{1}{2}, \tfrac{1}{2}\right) .$$

We denote that the characteristic of $\Omega_j$ , that is the set of vectors $a_j$ and $b_j$ , as $[\Omega_j]$. Also, the symbol $[\Omega_i]$, $[\Omega_{ij}]$, .... $[R]$ means their characteristics, a pair of vectors of dimension $g$ .

The characteristics $[\Omega_i]$, $[\Omega_{ij}]$, $[R]$ , are not always odd or not always even, as can be checked directly from (2.7), (2.8), (2.9). But if we do the following genus $g$ dependent procedures then we can say something more.

For example, in genus 2, if we add $R$ to $\Omega_i$ , $\Omega_{i,j}$ then the total of the characteristics will be as follows:

Characteristics $[\Omega_i + R]$, that is the pair of $a_i + \Delta^a$ *and* $b_i + \Delta^b$ : always odd for any value of $i$. (Total 6 cases, including $[R]$ itself which is $[\Omega_{2g+2} + R]$ )

Since $e_{2g+2} = e_6 = \infty$ , the value of any component in $[\Omega_6]$ is zero, so the characteristics are both zero vectors.

Characteristics $[\Omega_{ij} + R]$ : always even for all combinations of $i, j$ ( total $\binom{2g+1}{2} = 10$ cases.)

There are in total $2^{2g} = 16$ spin structures, $\frac{1}{2}(2^{2g} + 2^g) = 10$ are even and $\frac{1}{2}(2^{2g} - 2^g) = 6$ are odd.

In this sense, we say that $\Omega_i + R$ are non-singular and odd half-periods, and $\Omega_{ij} + R$ are non-singular even half periods at genus 2, although $[\Omega_i]$ is not always odd, and $[\Omega_{ij}]$ is not always even. For even characteristics $\delta$, the word non-singular means that the value of the theta function with zero characteristic at zero is not vanishing, $\theta_\delta(0) \neq 0$ .

We are only interested here in non-singular even half periods here, and so 10 $\Omega_{ij} + R$ are the $g = 2$ extension of half periods $\omega_\delta$, $\delta = 1,2,3$ in genus 1.

The above argument can be extended to any genus g in the hyperelliptic case. For the details, please see e.g. [11] .

It is known that in genus g, there are $\frac{1}{2}\binom{2g+2}{g+1}$ number of non-singular even half periods. This number is equal to the combinations of choosing $g$ number of $\Omega_i$ out of $2g+1$ number of $\Omega_i$ , $\binom{2g+1}{g}$ . In genus 3, the following results are known [13]:

$$
\begin{array}{lll}
[\,R\,] & 1 & \text{singular and even} \\
[\,\mathrm{R}+\Omega_i\,] & 2g+1=7 & \text{non singular and odd} \\
[\,\mathrm{R}+\Omega_{ij}] & \binom{2g+1}{2}=21 & \text{non singular and odd} \\
[\,R+\Omega_{ijk}\,] & \binom{2g+1}{3}=35 & \text{non singular and even}
\end{array}
\qquad (2.10)
$$

Total $2^{2g}=64$ spin structures, $\frac{1}{2}(2^{2g}+2^g)=36$ are even and $\frac{1}{2}(2^{2g}-2^g)=28$ are odd. Singular and even here in the first line above means that the zero characteristic theta function $\vartheta(R+v)$ vanishes at the origin $v=0$ ( to the order $m=2$ ) by Riemann's vanishing theorem.

Similarly, in genus 4, $\left[\,R+\Omega_{ijkl}\,\right]$ are non-singular and even spin structures [13].

We have summarized a description of the classification of all spin structures in genus $g$ for the hyper-elliptic case in Appendix C. Appendix C explains why both of $[\,\mathrm{R}+\Omega_i\,]$ and $[\,\mathrm{R}+\Omega_{ij}]$ in the case of genus 3 are non-singular and odd, for example.

**A proposition**

Suppose we choose any $g$ number of branch points $e_{i_1}, e_{i_2}, \dots e_{i_g}$ out of $2g+1$ branch points of arbitrary genus $g$ , with $e_{2g+2}$ fixed at $\infty$, and compute the integral (2.6). This determines one spin structure, and we denote $\delta$ :

$$\Omega_\delta=\Omega_{i_1 i_2 \dots i_g}=\left(\sum_{k=1}^{g} a_{i_k}\right)T+\left(\sum_{k=1}^{g} b_{i_k}\right) \quad . \qquad (2.11)$$

Then $[\,R+\Omega_\delta\,]$ will always be one of the characteristics of non-singular even spin structures by Riemann's theorem.

In the following, the theta function in genus $g$ whose characteristics are Riemann constants will be denoted as:

$$\theta_R(u) \stackrel{\text{def}}{=} \theta\begin{bmatrix}\Delta_a\\ \Delta_b\end{bmatrix}(u) \quad . \qquad (2.12)$$

The function $\theta_R(u)$ is odd if $\frac{g(g+1)}{2}$ is odd, and even if $\frac{g(g+1)}{2}$ is even, for the genus $g$.

The following proposition will be the starting point for the arguments on the decomposition formula in higher genus.

Consider a function of any genus in the hyperelliptic case defined as

$$F_g(u,\alpha) \stackrel{\text{def}}{=} \frac{\theta_R(u+\alpha)}{\theta_R(\alpha)E(z,w)} \tag{2.13}$$

where $u$ is the $g$-dimensional vector $u = (u_1, u_2, \dots u_g)$. The $\alpha$ is a $g$-dimensional parameter, $\alpha = (\alpha_1, \alpha_2, \dots \alpha_g) \in C^g$. $E(z,w)$ is the prime form, and set $u = A(z-w)$.

**Proposition 1**

Setting the parameter $\alpha$ as $\Omega_\delta$ where $\delta$ denotes a non-singular and even spin structure, we have, for an arbitrary genus $g$,

$$\frac{\theta[\delta](u)}{\theta[\delta](0)E(z,w)} = \exp\left(2\pi\mathrm{i}\left(\sum_{k=1}^{g} a_{i_k}\right)\cdot \mathrm{u}\right) \frac{\theta_R(u+\Omega_\delta)}{\theta_R(\Omega_\delta)E(z,w)} \quad . \tag{2.14}$$

Under the condition $z_{N+1} = z_1$, exp part becomes 1, so we have

$$\Pi_{i=1}^{N} S_\delta(z_i,\ z_{i+1}) = \Pi_{i=1}^{N}\ F_g(A(z_i - z_{i+1}),\ \Omega_\delta) = \Pi_{i=1}^{N} \frac{\theta_R(A(z_i - z_{i+1}) + \Omega_\delta)}{\theta_R(\ \Omega_\delta)E(z_i, z_{i+1})}\ . \tag{2.15}$$

Proof:

For any characteristic $A, B, C, D \in \frac{1}{2} Z^g$ all of which are $g$ component vectors and any variable $u$, we compare the two products $\frac{\theta\left[\begin{smallmatrix}A+C\\B+D\end{smallmatrix}\right](u)}{\theta\left[\begin{smallmatrix}A+C\\B+D\end{smallmatrix}\right](0)}$ and $\frac{\theta\left[\begin{smallmatrix}A\\B\end{smallmatrix}\right](u+TC+D)}{\theta\left[\begin{smallmatrix}A\\B\end{smallmatrix}\right](TC+D)}$.

Using

$$\theta\begin{bmatrix}A\\B\end{bmatrix}(u,T) = \theta(u+B+TA,\ T)\, exp\{\, i\pi A\cdot TA + 2\pi i A\cdot(u+B)\} \tag{2.16}$$

where $\theta(u+B+TA,\ T)$ is zero-characteristic theta function, we have

$$\frac{\theta\left[\begin{smallmatrix}A\\B\end{smallmatrix}\right](u+TC+D)}{\theta\left[\begin{smallmatrix}A\\B\end{smallmatrix}\right](TC+D)} = \exp(-2\pi\mathrm{i}\mathrm{C}\cdot\mathrm{u}) \frac{\theta\left[\begin{smallmatrix}A+C\\B+D\end{smallmatrix}\right](u)}{\theta\left[\begin{smallmatrix}A+C\\B+D\end{smallmatrix}\right](0)} \quad . \tag{2.17}$$

Therefore, if we set $\begin{bmatrix}A\\B\end{bmatrix}$ as $\begin{bmatrix}\Delta^a\\ \Delta^b\end{bmatrix}$, and $TC+D$ as $T(\sum_{k=1}^{g} a_{i_k}) + \sum_{k=1}^{g} b_{i_k}$, for a fixed non-singular even spin structure $\delta$, then we have

$$\frac{\theta_R(u+\Omega_\delta)}{\theta_R(\Omega_\delta)} = \exp\left(-2\pi\mathrm{i}\left(\sum_{k=1}^{g} a_{i_k}\right)\cdot\mathrm{u}\right)\ \frac{\theta[\delta](u)}{\theta[\delta](0)}. \tag{2.18}$$

Note that the pair of $A+C$ and $B+D$ is $\left[\ R+\ \Omega_{i_1 i_2 \dots i_g}\right]$.

Dividing both sides of (2.18) by the prime form $E(z,w)$ , we have eq. (2.14), (2.15).

As far as the fermion correlation functions are concerned, the function $F_g(u,\alpha)$ defined in (2.13) may be regarded as a generalisation of Eisenstein-Kronecker series of genus 1.

**On the second half of this paper**

We list up some facts and assumptions as (A) ~ (E) which will be helpful for further arguments below.

(A) is about the direction of the methods in the second half of this paper, and theorems (C) and (D) are important for our purpose. Although we will try to describe the cases of the general genus $g$, only the case of genus 2 is realistic when applied to the string theories, since in this paper all arguments are for hyper-elliptic cases.

Higher genus Pe functions are defined using theta functions with Riemann constant characteristics as:

$$P_{JK} = -\frac{\partial^2}{\partial u_J \partial u_K} ln\theta_R((2M)^{-1}u) - 2\eta(2M)^{-1}{}_{JK} \quad , \qquad P_{I_1 I_2 \dots I_N} = \frac{\partial^{N-2}}{\partial u_{I_3} \partial u_{I_4} \dots \partial u_{I_{N1}}} P_{I_1 I_2} \quad . \tag{2.19}$$

**(A) Expansions by higher genus Pe functions**

What we mainly consider is the product of $F_g(u_i,\alpha)$ which satisfies the cyclic condition $z_{i+1} = z_1$ :

$$\Pi_{i=1}^{N} F_g(u_i,\alpha) = \Pi_{i=1}^{N} \frac{\theta_R(A(z_i - z_{i+1}) + \alpha)}{\theta_R(\alpha) E(z_i, z_{i+1})} \quad . \tag{2.20}$$

We assume that the right-hand side of (2.20) can be expanded by a set of base functions, as in the case of genus 1. As already described, the Laurent expansion form of $ln\theta_R(u)$ is not known for $g > 1$. Because of this fact, we emphasize that in the second half of this paper we have to leave many results unproved. What we can do is to ask what we would expect to see in the spin sum if the natural analogy of the case of genus 1 were to hold for the case of higher genus.

**(B) Basis functions of the expansion**

The base functions to expand the right-hand side of (2.20) will be spin structure dependent factors by setting the g-dimensional parameter $\alpha$ equal to $\Omega_\delta$ where $\delta$ represents non-singular and even spin structures.

By reflecting the cyclic condition, as in the case of genus 1, such base functions will be expressed by series of higher genus Pe functions, since the exp factor in (2.18) falls in the product (2.20). The most natural candidates of such base functions will be $P_{IJ}(\alpha)$, $P_{IJK}(\alpha)$, $P_{IJKL}(\alpha)$, , $P_{IJKLM}(\alpha)\ldots$ , but it will turn out that the dimension of the functional space of these is not large enough to derive correct results for higher genus cases. ( See section 2.4, below of eq. (2.68) )
At the level of genus one, there is another set of base functions

$P(\alpha),\ P^{(1)}(\alpha), P^2(\alpha),\ P^{(1)}(\alpha)\,P(\alpha),\ P^3(\alpha),\ P^{(1)}(\alpha)P^2(\alpha)\ ,\ P^4(\alpha)\ ,\ P^{(1)}(\alpha)P^3(\alpha),\ \ldots$

which are equivalent to those used in section one, and we assume that a generalization of these would give the correct expansion base functions.

We prepare all such base functions with appropriate order of poles, such as

$$P_{AB}(\alpha),\ P_{ABC}(\alpha),\ P_{AB}(\alpha)P_{CD}(\alpha),\ P_{ABC}(\alpha)P_{DE}(\alpha),\ P_{AB}(\alpha)\,P_{CD}(\alpha)P_{EF}(\alpha)\ldots\ldots \qquad (2.21)$$

For the even order of poles, we prepare the product of $P_{AB}(\alpha)$ s. For the odd order of the poles[2] ,we multiply the odd order pole functions, $P_{ABC}(\alpha), P_{ABCDE}(\alpha)$ , … by the product of $P_{AB}(\alpha)$ s.

It seems that the following proposition can be validated for a non-singular even spin structure $\Omega_\delta$ :

$$P_{I_1I_2\ldots I_{2k-1}}(\Omega_\delta)=0 \quad \text{if } k \text{ is any positive integer .} \qquad (2.22)$$

as a generalised formula of $P^{(ODD)}(\omega_\delta)=0$ in genus one (see Appendix D-2). Therefore, after setting $\alpha=\Omega_\delta$ , only the terms of simple products of $P_{AB}$, $P_{AB}P_{CD}$, $P_{AB}\,P_{CD}P_{EF}$,.. ( without odd order $P_{ABC}(\alpha), P_{ABCDE}(\alpha)$ , …) will remain non-zero.

From the form of the prime forms in the product of the fermion correlation function of the left-hand side of (2.20) under the condition $z_{N+1}=z_1$, the whole of the product of eq. (2.20) will be holomorphic 1 forms, and the $z_i$ dependence of $\omega_{I_1}$, $\omega_{I_2}$, ... $\omega_{I_N}$ will be all different. This is also supported by the results of ref. [7][8]. That is, one may have to search for a possible form as

$$\Pi_{i=1}^{N}\ F_g(A(z_i-z_{i+1}),\alpha)=J_{N,N}^{I_1I_2\ldots I_N}\omega_{I_1}(z_1)\omega_{I_2}(z_2)\ldots\omega_{I_N}(z_N)$$
$$+\ J_{N,N-2}^{I_1I_2\ldots I_N}\omega_{I_1}(z_1)\omega_{I_2}(z_2)\ldots\omega_{I_N}(z_N)\ \ldots+J_{N,0}^{I_1I_2\ldots I_N}\omega_{I_1}(z_1)\omega_{I_2}(z_2)\ldots\omega_{I_N}(z_N) \qquad (2.23)$$

where $J_{N,N}^{I_1I_2\ldots I_N}$ does not depend on the Pe function, $J_{N,N-2}^{I_1I_2\ldots I_N}$ depends on $P_{I_1I_2}$, $J_{N,N-3}^{I_1I_2\ldots I_N}$ depends on $P_{I_1I_2I_3}$, …., . Implicit sums over $I_1I_2\ldots I_N=1,2,\ldots g$ are assumed. In higher genus, the notation of $J_{N,M}^{I_1I_2\ldots I_N}$ is used instead of $H$, because of the intricate contractions of indices between the modular covariant function parts and the Pe functions parts.

---

[2] Here the word "order of poles" is used in the sense as described in the footnote 6 of page 32.

Basically, we are interested in the pole structures on the difference of the vertex insertion points $z_i - z_{i+1}$ . The product of fermion correlation functions should be the one that can reproduce "simultaneously single poles of the products of $z_i - z_{i+1}$ for different $i$ " , as in the case of genus one. In the following, however, only the differentiations with respect to $u_J$ defined as $u = (u_1, u_2, \dots u_g)$ with

$$u_1 = \left(\int_{z_i}^{z_{i+1}} \omega_1\right), \ \ .. \ u_g = \left(\int_{z_i}^{z_{i+1}} \omega_g\right)$$

will appear. These $u_J$ are the variables in the theta functions of higher genus. We denote these differentiations $\frac{\partial}{\partial u_J}$ as $\frac{\partial}{\partial_J}$ or $\partial_J$ . We will look for possible forms of $J_{N,M}^{I_1 I_2 \dots I_N}$ containing only differentiations $\partial_J$, whose index $J$ is contracted with the 1 forms $\omega_{I_1}(z_1)\omega_{I_2}(z_2) \dots \omega_{I_N}(z_N)$ . We expect that, if we choose appropriate forms of $J_{N,M}^{I_1 I_2 \dots I_N}$ , then the desired poles of $z_i - z_{i+1}$ will naturally be reproduced in a way that we require as the decomposition formula of the product of fermion correlation functions, as a generalisation of genus one results.

If we admit (A), (B), then, as in the case of g=1, the spin structure dependence of the product (2.20) with the condition $z_{i+1} = z_1$ is entirely via $P_{IJ}(\Omega_\delta)$ which is a symmetric $g \times g$ matrix.

The following (C) is a known fact derived from equations in solutions of Jacobi's inversion problem. It is a generalization of $e_\delta = P(\omega_\delta)$ in genus one.

(C) There is a general method to determine all of $\frac{g(g+1)}{2}$ number of elements of a symmetric matrix $P_{IJ}(\Omega_\delta)$ by solving the $\frac{g(g+1)}{2}$ number of equations related to the solutions of Jacobi's inversion problem. All of elements of $P_{IJ}(\Omega_\delta)$ can be expressed by branch points $e_1, e_2, \dots\dots e_{2g+1}$ for any genus $g$ in the hyperelliptic case.

For example, in genus 2, for any fixed spin structure $\Omega_\delta = \Omega_{ij}$ ,

$$P_{22}(\Omega_{ij}) = e_i + e_j, \quad P_{12}(\Omega_{ij}) = P_{21}(\Omega_{ij}) = -e_i e_j \, , \tag{2.24}$$

$$P_{11}(\Omega_{ij}) \ = \ (e_p + e_q + e_r)e_i e_j + e_p e_q e_r \tag{2.25}$$

where each of $e_p, \ e_q, \ e_r$ is different from $e_i, \ e_j$ .

The modular weights of $P_{22}(\Omega_{ij}), \ P_{12}(\Omega_{ij}), \ P_{11}(\Omega_{ij})$ are different from each other.

In arbitrary genus in hyperelliptic cases, for fixed values of $J, K,$ the expression of $P_{JK}(\Omega_{i_1 i_2 i_3 \dots i_g})$ depends on $e_{i_1}, \ e_{i_2}, \dots\dots e_{i_g}$ which are one set of choice of $g$ number of branch points out of $2g+1$ branch points. This set determines one non-singular even spin structure $\delta$.

(D) It holds that, as the generalisation of Weierstrass formula of genus one eq.(1.46), in the arbitrary genus of the hyperelliptic case,

$$[2\eta(2M)^{-1}]_{IJ} = -P_{IJ}(\Omega_\delta) - [\,((2M)^{-1})^T H (2M)^{-1}]_{IJ} \qquad (2.26)$$

where a matrix $H$ is defined as

$$H_{KL} = \frac{\partial_K \partial_L\, \theta[\delta](0)}{\theta[\delta](0)} \qquad (2.27)$$

for $I, J = 1,2, \dots g$ . (See eq. (2.100) in [10]. The original paper is [12]. See also Appendix D-2. ) The $\delta$ represents any of non-singular even spin structures.

We expect that the following (E) may also be validated in future.

(E) After summation over non-singular even spin structures, which can be done by purely algebraic calculations of the branch points $e_1, e_2, \dots\dots e_{2g+1}$ , the result will be represented by symmetric functions of $2g+1$ branch points. All fundamental symmetric functions of $2g+1$ branch points may be modular covariant forms which are generalisations of holomorphic Eisenstein series in genus 1.

The function (2.20) has the same form as in the case of genus 1, (1.22), (1.23). In particular, there is a factor $\theta_R(\alpha)$ in the denominator of the product. Therefore, the same argument applies in the case of genus one, on pages 8 and 9. That is, the "zero mode subtraction" procedure which we saw in the case of genus 1 may also hold for the same reason.

In order to pursue the similarity with the case of genus one, we also assume this property as follows[3].

We define the zero mode as

$$\Lambda_{I_1 I_2 \dots I_M} \stackrel{\text{def}}{=} [\partial^M_{I_1 I_2 \dots I_M} \ln \theta_R(\alpha)\,]_{\alpha=0} \qquad (2.28)$$

where $\ln \theta_R(\alpha)$ is a theta function a genus $g$ with Riemann constants characteristic which has an arbitrary variable $\alpha$ in the hyperelliptic case. The $\alpha$ is a $g$ dimensional vector whose components are $\alpha_1, \alpha_2, \dots \alpha_g$ and $[\ \ ]_{\alpha=0}$ means to extract the constant term, $(\alpha_1\, \alpha_2 \dots \alpha_g)^0$ term, from the non-singular (at the origin) terms.

$\partial^M_{I_1 I_2 \dots I_M}$ means that differentiating with respect to $\alpha_{I_1}, \alpha_{I_2}, \dots \alpha_{I_g}$ a total $M$ times for each of $I_k$ taking values either of $1,2, \dots g$ .

The set of all of "zero modes" $\Lambda_{I_1 I_2 \dots I_M}$ is the same as the set of all expansion coefficients of $\ln\theta_R(\alpha)$ , which may be the set of generalised holomorphic Eisenstein series in higher genus, since $\ln\theta_R(\alpha)$ is the function also used in the definition of the higher genus Pe function as in (2.19).

---

[3] To explain the concept, we first describe the case where $2M_{IJ}$ is normalised, $2M_{IJ} = \delta_{IJ}$ . The real calculations are described in 2.2.

We also define

$$D^M_{I_1 I_2 \dots I_M} \ln\theta_R(u) \ \stackrel{\text{def}}{=}\ \partial^M_{I_1 I_2 \dots I_M} \ln\theta_R(u) \ - \ \Lambda_{I_1 I_2 \dots I_M} \quad . \qquad (2.29)$$

We define the following $W_{I_1,I_2,\dots I_K}(z_1, z_2, \dots . z_N)$ which is a generalisation of the function $V$ in the genus one case:

$$W_{I_1,I_2,\dots I_K}(z_1, z_2, \dots . z_N)$$

$$\stackrel{\text{def}}{=}\ D^K_{I_1 I_2 \dots I_K} \exp\left( \sum_{i=1}^{N} \ln\theta_R(A(z_i - z_{i+1}) + \alpha) - \ \sum_{i=1}^{N} \ln\theta_R(A(z_i - z_{i+1})\,)\right)\Big|_{\alpha=0,} \qquad (2.30)$$

After performing all differentiations $\partial^K_{I_1 I_2 \dots I_K}$ and setting $\alpha = 0$, all results of $\partial^K_{I_1 I_2 \dots I_K} \ln\theta_R(u)$ should be replaced with $\partial^K_{I_1 I_2 \dots I_K} \ln\theta_R(u) \ - \ \Lambda_{I_1 I_2 \dots I_K}$. By setting $\alpha = 0$ , each of the differentiations of the theta function $\ln\theta_R(u)$ $(u_i = A(z_i - z_{i+1}\,))$ is with respect to $u_i$

At this stage, where the expansion form of $\ln\theta_R(u)$ is not known, it is difficult to pursue rigorous mathematical arguments further. In what follows, we adopt one more procedure. We restrict our argument to the case that $\theta_R(u)$ is odd, i.e. the genus g satisfies $\frac{g(g+1)}{2}$ is odd, and adopt the odd theta function in the prime form as $\theta_R(A(z_i - z_{i+1}))$ itself.

When $\theta_R(u)$ is even, the coefficients of the expansion form of $ln\theta_R(u)$ may have different properties from those in the case of odd $\theta_R(u)$, and the procedure (A) may not work well. In the case of even $\theta_R(u)$ , the author feels that different kind of consideration may be needed from the beginning.

## 2.2 Candidates for a decomposition formula for small N

In this subsection, we describe some calculation examples assuming that the expansion functions of the product (2.20) are $P_{IJ}(\alpha)$, $P_{IJK}(\alpha)$, $P_{IJKL}(\alpha)$, , $P_{IJKLM}(\alpha) \dots$ , just Pe function and its derivatives, as a direct generalisation of genus 1 case. This set is different from those given in (B) or in Appendix B, the space is smaller. As we will see later, the set used in this subsection does not have enough functional space to obtain correct expansion coefficients in general values of $N$. However, for small $N$, the results of the decomposition formula become the same as the correct ones, and it may give an example of calculations because we can borrow the same type of polynomials of genus 1 case rigorously obtained in Chapter 1.

**Case N=2**

N=2 $S_\delta(z_1,z_2)\, S_\delta(z_2,z_1)$

We set the product $\Pi_{i=1}^{2}\ F_g(A(z_i - z_{i+1}),\alpha)$ under the condition $z_3 = z_1$ as

$$\Pi_{i=1}^{2}\ F_g(A(z_i - z_{i+1}),\alpha)$$

$$= \sum_{I,J=1,2,..g} J_{2,2}^{IJ}\omega_I(z_1)\omega_J(z_2) \quad + \quad \sum_{I,J=1,2,..g} J_{2,0}^{IJ}\omega_I(z_1)\omega_J(z_2) \qquad (2.31)$$

where $J_{2,2}^{IJ}$ is $P_{IJ}(\alpha)$ independent term, and $J_{2,0}^{IJ}$ is the Pe function dependent term. We try to find the coefficients $J_{2,2}^{IJ}$ and $J_{2,0}^{IJ}$ by analogy with the genus one case, eq. (1.42).

$J_{2,0}^{IJ}$ will be $P_{IJ}(\alpha)$ ,by looking at eq. (1.42) in genus one.

As for $J_{2,2}^{IJ}$ , the form will be, by looking at $V_2(x_1,x_2)$ in (1.43),

$$J_{2,2}^{IJ} = D_{IJ}^2\, ln\, \theta_R\, ((2M)^{-1}A(z_1 - z_2)) \qquad (2.32)$$

In $D_{IJ}^2$ , the first "zero mode" $\Lambda_{IJ}$ would be the second period matrix $2\eta_{IJ}$ in (2.4), which is the analogue of $2\eta_1$ or $G_2$ in the genus 1. That is,

$J_{2,2}^{IJ}$ will be $-P_{IJ}((2M)^{-1}A(z_1 - z_2))$ .

If we set $\alpha = \Omega_\delta$ , eq. (2.31) becomes

$$-S_\delta(z_1,z_2)^2 = \sum_{I,J}\left[\partial_I\partial_J\ ln\,\theta_R\left((2M)^{-1}A(z_1 - z_2)\right) + [2\eta(2M)^{-1}]_{IJ}\right]\omega_I(z_1)\omega_J(z_2) + $$
$$+ \sum_{I,J} P_{IJ}(\Omega_\delta)\omega_I(z_1)\omega_J(z_2)\ . \qquad (2.33)$$

We use the above theorem (D) here：

$$S_\delta(z_1,z_2)^2 = -\sum_{I,J}\partial_I\partial_J\ ln\,\theta_R\left((2M)^{-1}A(z_1 - z_2)\right)\omega_I(z_1)\omega_J(z_2) + $$
$$+ \sum_{I,J}\omega_I(z_1)\omega_J(z_2)\,[\,((2M)^{-1})^T H (2M)^{-1}]_{IJ} \qquad (2.34)$$

On the other hand, we know an exact formula for the square of the fermion correlation functions [7], in the case that $(2M)$ is normalized as $\delta_{IJ}$ ：

$$S_\delta(z_1,z_2)^2 = \partial_{z_1}\partial_{z_2} lnE(z_1,z_2) + \sum_{I,J}\omega_I\,(z_1)\ \omega_J(z_2)\frac{\partial^I\partial^J\ \theta[\delta](0)}{\theta[\delta](0)} \qquad (2.35)$$

In the prime form $E(z,w) = \frac{\theta[\nu](A(z-w))}{h_\nu(z)h_\nu(w)}$ , each of $h_\nu(z)$, $h_\nu(w)$ depends on only one variable, $z_1$ or $z_2$ . Then $\partial_{z_1}\partial_{z_2}\ln\,(h_\nu(z_1)h_\nu(z_2)) = 0$ , and (2.35) becomes

$$S_\delta(z_1,z_2)^2 = \partial_{z_1}\partial_{z_2} ln\theta[\nu]\left(A(z_1 - z_2)\right) + \sum_{I,J}\omega_I\,(z_1)\ \omega_J(z_2)\frac{\partial^I\partial^J\ \theta[\delta](0)}{\theta[\delta](0)} \qquad (2.36)$$

for an odd theta function $\theta[\nu]\left(A(z_1 - z_2)\right)$ .

Using the chain rules of differentiation, (2.36) has the same form as (2.34).

The formula (D) (Weierstrass formula in genus g) says that the non-singular even spin structure dependence of the product of fermion correlation functions with cyclic condition should be through only one kind of theta constants, $\frac{\partial^I \partial^J \theta[\delta](0)}{\theta[\delta](0)}$ for arbitrary $N$ , if the spin structure depends on only one kind of constants $P_{IJ}(\Omega_\delta)$ , analogous to the case in genus 1.

**Case N=3**

N=3 $S_\delta(z_1, z_2)\, S_\delta(z_2, z_3) S_\delta(z_3, z_1)$

$$\Pi_{i=1}^{3} F_g(A(z_i - z_{i+1}), \alpha) = \sum_{I,J,K=1,2,..g} J_{3,3}^{IJK}\, \omega_I(z_1)\omega_J(z_2)\, \omega_K(z_3)$$
$$+ \sum_{I,J,K=1,2,..g} J_{3,1}^{IJK}\, \omega_I(z_1)\omega_J(z_2)\omega_K(z_3) + \sum_{I,J,K=1,2,..g} J_{3,0}^{IJK}\, \omega_I(z_1)\omega_J(z_2)\omega_K(z_3) \quad (2.37)$$

A natural candidate of the form of $J_{3,i}^{IJK}$ may be constructed as follows.

Define $L_N$ as

$$L_N \stackrel{\text{def}}{=} \sum_{i=1}^{N} ln\, \theta_R\left((2M)^{-1} A(z_i - z_{i+1})\right) \ . \quad (2.38)$$

Considering the one loop case (1.49), $J_{3,3}^{IJK}$ may have the form

$$J_{3,3}^{IJK} = \frac{1}{6}\, \partial_{IJK}\, L_3 + \frac{1}{6} \{ \, [\partial_I\, L_3][\, D_{JK}^2\, L_3] + [\partial_J\, L_3][\, D_{IK}^2\, L_3] + [\partial_K\, L_3][\, D_{IJ}^2\, L_3] \, \}$$
$$+ \frac{1}{6} [\, \partial_I\, L_3][\, \partial_J\, L_3][\, \partial_K\, L_3] \quad (2.39)$$

The meaning of the differentiations $\partial_I$ is as follows. Define $u_i \stackrel{\text{def}}{=} A(z_i - z_{i+1})$

Each $u_i$ is a vector of dimension $g$ , $u_i^I = \left( \int_{z_i}^{z_{i+1}} \omega_I \right)$ The meaning of $\partial_I L_N$ is that for each term of $ln\, \theta_R\left((2M)^{-1} u_i\right)$ within the sum of $L_N$, $ln\, \theta_R\left((2M)^{-1} u_i\right)$ is differentiated with respect to $u_i^I$ , and the index $I$ is contracted with $\omega_I$ .

Since we use the notation (2.3), the result of $\partial_I L_3$ should be written as $(2M)_{Ij}^{-1} \partial_j L_3((2M)^{-1} u\ \ )_j)$ as we saw in $H$ matrix in (2.26), using a matrix $(2M)^{-1}$ .
We here assume in the below that it is agreed that any types of the derivatives of $L_N$ shall be written as $\partial_I L_N$ etc., abbreviating such complicated forms using $(2M)^{-1}$.

The middle terms in the right-hand side of (2.39) are further divided by 3 because they contain the cyclic sum of $I, J, K$, compared to the genus one case of Eq. (1.49).

We also have

$$J_{3,1}^{IJK} = \frac{1}{3} \{ \, [\partial_I\, L_3][\, P_{JK}(\alpha)] + [\partial_J\, L_3][P_{IK}(\alpha)\,] + [\partial_K\, L_3][\, P_{IJ}(\alpha)] \, \} \quad (2.40)$$
$$J_{3,0}^{IJK} = \frac{1}{6} P_{IJK}(\alpha) \quad (2.41)$$

Setting $\alpha = \Omega_\delta$, since $P_{IJK}(\Omega_\delta) = 0$, we have[4], assuming the implicit summation over $I, J, K$ =1,2,..g ,

$$S_\delta(z_1,z_2)S_\delta(z_2,z_3)S_\delta(z_3,z_1) = J_{3,3}^{IJK}\,\omega_I(z_1)\omega_J(z_2)\omega_K(z_3)$$
$$+\frac{1}{3}\{\ [\partial_I\, L_3]\big[P_{JK}(\Omega_\delta)\big] + [\partial_J\, L_3][P_{IK}(\Omega_\delta)\,] + [\partial_K\, L_3]\big[P_{IJ}(\Omega_\delta)\big]\ \}\,\omega_I(z_1)\omega_J(z_2)\omega_K(z_3) \qquad (2.42)$$

**Case N=4**

N=4 $\quad S_\delta(z_1,z_2)S_\delta(z_2,z_3)S_\delta(z_3,z_4)S_\delta(z_4,z_1)$

We write the form

$$\Pi_{i=1}^{4}\ F_g(A(z_i - z_{i+1}),\alpha) = J_{4,4}^{IJKM}\,\omega_I(z_1)\omega_J(z_2)\ \omega_K(z_3)\ \omega_M(z_4)$$
$$+J_{4,2}^{IJKM}\,\omega_I(z_1)\omega_J(z_2)\ \omega_K(z_3)\ \omega_M(z_4) + \ J_{4,1}^{IJKM}\,\omega_I(z_1)\omega_J(z_2)\ \omega_K(z_3)\ \omega_M(z_4)$$
$$+J_{4,0}^{IJKM}\,\omega_I(z_1)\omega_J(z_2)\ \omega_K(z_3)\ \omega_M(z_4) \qquad (2.43)$$

[ Note: As described at the beginning of subsection 2.2,
the set of functions $P_{IJ}(\alpha)$, $P_{IJK}(\alpha)$, $P_{IJKL}(\alpha)$, $P_{IJKLM}(\alpha)$ ... has too small function space dimension to give correct results for large values of $N$. For $N = 4$, the result $J_{4,0}^{IJKM}$ given in (2.47) below does not give the correct spin sum result obtained in [7]. Please see (2.66) and below. ]

Looking at the genus one $V_4(x_1, x_2, \dots . x_4)$ in (1.55), $J_{4,4}^{IJKM}$ may have the form

$$J_{4,4}^{IJKM} = \frac{1}{24}[\,\partial_I\, L_4][\,\partial_J\, L_4][\,\partial_K\, L_4][\,\partial_K\, L_4]$$
$$+\frac{1}{24}\{\{[\,\partial_I\, L_4][\,\partial_J\, L_4] + [\,D_{IJ}^2\, L_4]\}[\,D_{KM}^2\, L_4] + (cycl\ I,J,K,M)\,\}$$
$$+\frac{1}{24}\{\ [\,\partial_I\, L_4]\big[\,\partial_{JKM}^3\, L_4\big] + (cycl\ I,J,K,M)\}\ + \frac{1}{24}\,D_{IJKM}\,L_4 + \ \ f(\,\Lambda_{IJKM}) \qquad (2.44)$$

The terms in the second row of (2.44) are divided by 6 which is the number of selections of two indices $I, J$ out of four indices, and the third terms are divided by 4 which is the number of cyclic selections of one index $I$ out of 4. As a result, all the numerical numbers become $\frac{1}{24}$. This is not surprising, since it is equivalent to the result of differentiating eq. (2.30) with respect to $\alpha_1$ and $\alpha_2$ four times using different indices and then setting both to zero, as if (2.30) were the generating function of the modular covariant part of the decomposition formula.

---

[4] If we use the modified form of $V_3$ (1.52) instead of (1.49), the result of $J_{3,3}^{IJK}$ will be

$$J_{3,3}^{IJK} = \frac{1}{6}\,\partial_{IJK}\,L_3 + \frac{1}{9}\{\ [\,\partial_I\, L_3]\big[\,D_{JK}^2\, L_3\big] + [\,\partial_J\, L_3][\,D_{IK}^2\, L_3] + \ [\,\partial_K\, L_3]\big[\,D_{IJ}^2\, L_3\big]\ \}$$

This matches with the calculations in ref. [2] for the $N = 3$ case of genus 2. Calculations staring from Fay's formula often go to get such modified expressions.

This $\frac{1}{24}$ in (2.44), or $\frac{1}{6}$ for the $N=3$ case in (2.39), obviously comes from the $\frac{1}{k!}$ factor in eq. (1.34) which is for the genus one case where there is only one scalar parameter $\alpha$ to consider. Since we are doing a multivariable expansion of $\alpha_1, \alpha_2, \dots \alpha_g$ , these numerical factors will be modified in rigorous arguments.

For other terms, we have

$$J^{IJKM}_{4,2} = \frac{1}{12}\{\{[\partial_I L_4][\partial_J L_4] + [D^2_{IJ} L_4]\}[P_{KM}(\alpha)] + (cycl\ I,J,K,M)\} \tag{2.45}$$

$$J^{IJKM}_{4,1} = \frac{1}{4}\{[\partial_I L_4][P_{JKM}(\alpha)] + (cycl\ I,J,K,M)\} \tag{2.46}$$

$$J^{IJKM}_{4,0} = \frac{1}{6} P_{IJKM}(\alpha) \tag{2.47}$$

Setting $\alpha = \Omega_\delta$, since $P_{JKM}(\Omega_\delta) = 0$ , we have

$$S_\delta(z_1,z_2)S_\delta(z_2,z_3)S_\delta(z_3,z_4)S_\delta(z_4,z_1) = J^{IJKM}_{4,4}\,\omega_I(z_1)\omega_J(z_2)\omega_K(z_3)\omega_M(z_4) +$$
$$\frac{1}{12}\{\{[\partial_I L_4][\partial_J L_4] + [D^2_{IJ} L_4]\}[P_{KM}(\Omega_\delta)] + (cycl\ I,J,K,L)\}\,\omega_I(z_1)\omega_J(z_2)\omega_K(z_3)\omega_M(z_4) +$$
$$\frac{1}{6} P_{IJKM}(\Omega_\delta)\,\omega_I(z_1)\omega_J(z_2)\omega_K(z_3)\omega_M(z_4) \tag{2.48}$$

There will be another kind of numerical factors to consider. In genus one, the origin of the factor $\frac{1}{(2K-1)!}$ in front of eq. (1.12) was naturally explained in the proof of the decomposition formula. This factor was first correctly guessed in [3]. The similar overall factor will also appear in the higher genus case and we may have to guess it. This is not a trivial problem and I have no idea about it at the moment. The difficulty comes from the fact that it is not so easy to express $P_{I_1 I_2 \dots I_M}(\alpha)$ ( $M$ $even$) by the polynomial of $P_{I_1 I_2}(\alpha)$ for arbitrary even values of $M$.

In $J^{IJKM}_{4,4}$, "next zero mode", $\Lambda_{IJKM}$, which corresponds to the Eisenstein series $G_4$ in the case of genus 1 case, may appear in $\frac{1}{24} D_{IJKM} L_4$ , whose concrete form we don't know yet. This second zero mode may be cancelled by the additional terms in $J_{N,N}$ , something like a generalised form of $-\sum_{K=2}^{[\frac{N}{2}]} G_{2K} \cdot V_{N-2K}$ in the genus one case. It would be difficult to find out the exact form of these additional terms of $J_{N,N}$ without an exact proof of the whole of the decomposition formula. As we saw in the case of genus one, it is not possible to determine these by the pole structure considerations.

## 2.3 Zero modes from the base functions of expansion

In the last subsection 2.2, we described calculation examples assuming that the expansion functions of the product (2.20) are

$$P_{IJ}(\alpha),\ P_{IJK}(\alpha),\ P_{IJKL}(\alpha),,\ P_{IJKLM}(\alpha) \dots, \tag{2.49}$$

as a direct generalisation of the case of genus 1. Up to $N = 3$, this may give correct form of the decomposition formula, but as we will see later, we need to use more general base functions to expand the product (2.20) for $N \geq 4$ in higher genus $g \geq 2$.
In this subsection, we will digress a bit and describe what form the expansion of (2.20) would look like if it were expanded by (2.49) in general.

In such a case, by considering how the "additional terms" $-\sum_{K=2}^{\left[\frac{N}{2}\right]} G_{2K} \cdot V_{N-2K}$ come from $H_{N,N}$ in the proof given in Chapter 1, the product of the fermion correlation functions will have the following closed form:

$$\Pi_{i=1}^{N} S_\delta(z_i, z_{i+1}) = W_{I_1 I_2 \dots I_N} \ \omega_{I_1}(z_1)\omega_{I_2}(z_2) \dots \omega_{I_N}(z_N)$$
$$+ \sum_{K=1}^{\left[\frac{N}{2}\right]} \left[ W_{I_1 I_2 \dots I_{N-2K}} \ P_{I_{N-2K+1} I_{N-2K+2} \dots I_N}(\Omega_\delta) + (cyclic \ \ I_1, I_2, .. I_N) \right] \omega_{I_1}(z_1)\omega_{I_2}(z_2) \dots \omega_{I_N}(z_N)$$
$$+ \sum_{K=2}^{\left[\frac{N}{2}\right]} \left[ W_{I_1 I_2 \dots I_{N-2K}} \ \Lambda_{I_{N-2K+1} I_{N-2K+2} \dots I_N} + (cyclic \ \ I_1, I_2, .. I_N) \right] \omega_{I_1}(z_1)\omega_{I_2}(z_2) \dots \omega_{I_N}(z_N)$$

(2.50)

In the last line, the sum starts from $K = 2$ because the first zero mode lambda $\Lambda_{IJ}$, which will be $-2\eta_{IJ}$ for any genus, is already included in the Pe function in the second line.

The first line and the third lines give terms that do not depend on the spin structures, as a generalisation of $V_N - \sum_{K=2}^{\left[\frac{N}{2}\right]} G_{2K} \cdot V_{N-2K}$ in the case of genus one.
This can also be written as

$$\Pi_{i=1}^{N} S_\delta(z_i, z_{i+1}) = W_{I_1 I_2 \dots I_N} \ \omega_{I_1}(z_1)\omega_{I_2}(z_2) \dots \omega_{I_N}(z_N)$$
$$- \sum_{K=1}^{\left[\frac{N}{2}\right]} \left[ W_{I_1 I_2 \dots I_{N-2K}} \ D^{2K}_{I_{N-2K+1} \dots I_N} \ln \theta_R(\Omega_\delta) + (cyclic \ \ I_1, I_2, .. I_N) \right] \omega_{I_1}(z_1)\omega_{I_2}(z_2) \dots \omega_{I_N}(z_N)$$

(2.51)

where

$$D^{2K}_{I_{N-2K+1} I_{N-2K+2} \dots I_N} \ln \theta_R(\Omega_\delta) = \partial^{2K}_{I_{N-2K+1} \dots I_N} \ln \theta_R(\Omega_\delta) - \Lambda_{I_{N-2K+1} I_{N-2K+2} \dots I_N}.$$

In the case of genus 1, we can apply (2.51) rigorously. we can write the formula (1.11) as

$$\Pi_{i=1}^{N} S_\delta(z_i - z_{i+1}) \; = \; V_N - \sum_{K=1}^{\left[\frac{N}{2}\right]} \frac{1}{(2K-1)!} V_{N-2K} \cdot D^{2K} \ln \theta_1(\omega_\delta) \tag{2.52}$$

where[5]

$$D^{2K} \ln \theta_1 (\omega_\delta) \stackrel{\text{def}}{=} \frac{\partial^{2K}}{\partial \alpha^{2K}} \ln \theta_1 (\alpha \;\;, \tau)\bigg|_{\alpha=\omega_\delta} + (2K-1)!\, G_{2K} \tag{2.53}$$

The genus 1 zero modes are $-(2K-1)!\, G_{2K}$. The zero-mode subtraction procedure occurs twice in this formula, once is in the $V$, and once in $D^{2K} \ln \theta_1 (\omega_\delta)$.

As explained, the zero mode subtraction in $V$ results from the existence of $\theta_1(\alpha, \tau)$ in the denominator of the product $\Pi_{i=1}^{N} F(x_i, \alpha, \tau) = \Pi_{i=1}^{N} \frac{\theta_1^{(1)}(0)\theta_1(x_i+\alpha,\tau)}{\theta_1(x_i,\tau)\theta_1(\alpha,\tau)}$ .

On the other hand, zero mode subtraction in $D^{2K} \ln \theta_1 (\omega_\delta)$ originates from the zero modes ($\alpha^0$ terms) in the set of expansion base functions $P$ , $P^{(1)}$, $P^{(2)}$ .... .

The coefficients $G_{2K}$ of $\sum_{K=2}^{\left[\frac{N}{2}\right]} G_{2K} \cdot V_{N-2K}$ in $H_{N,N}$ have such an origin. The formula (2.50) or (2.51) can't be applied directly to higher genus, but it would be quite plausible that in the decomposition formula spin structure independent terms (generalisation of $H_{N,N}$ ) will include zero modes of expansion base functions. To determine those terms we need a rigorous derivation because the pole structure investigation does not work as we saw in the genus 1 case.

## 2.4 Consistency checks and dimension of space of expansion base functions

**Genus 2 case , $(N = 2) \times (N = 2)$ product**

For the genus 2, N=4 point amplitudes calculations of the paper [7] , the following spin sum was performed:

$$I_{20}(z_1, z_2, z_3, z_4) = \sum_\delta \Xi_6[\delta]\vartheta[\delta](0)^4 S_\delta(z_1, z_2)^2 S_\delta(z_3, z_4)^2$$

$$= \; -4\pi^4 \Psi_{10}(\Delta(z_1, z_3)\Delta(z_2, z_4) + \Delta(z_1, z_4)\Delta(z_2, z_3)\,) \tag{2.54}$$

$$\Delta(x, y) = \omega_1(x)\omega_2(y) - \omega_2(x)\omega_1(y) \tag{2.55}$$

$I_{20}$ corresponds to the $(N = 2) \times (N = 2)$ product of the fermion correlation functions. In our method, the only terms that can be non-zero after the spin sum are

$$[\, \sum_{I,J} P_{IJ}(\Omega_\delta)\omega_I(z_1)\omega_J(z_2)][\, \sum_{K,L} P_{KL}(\Omega_\delta)\omega_K(z_3)\omega_L(z_4)] \; . \tag{2.56}$$

In genus 2, if the string partition function is expressed by 5 branch points and summed over, this should give $\Delta(z_1, z_3)\Delta(z_2, z_4) + \Delta(z_1, z_4)\Delta(z_2, z_3)$ in (2.54).

---

[5] The minus sign in front of the second term of the right hand side of (2.52) comes from the definition of Pe function, (A.8).

The descriptions of the spin sum results in ref. [2] page 37 for genus 2 are:

$$\sum_\delta P_{12}P_{12} = 2\sqrt{D} \qquad \sum_\delta P_{11}P_{22} = -4\sqrt{D} \tag{2.57}$$

$$\sum_\delta P_{12}P_{11} = \sum_\delta P_{11}P_{11} = \sum_\delta P_{22}P_{22} = \sum_\delta P_{22}P_{12} = 0 \tag{2.58}$$

Eqs.(2.57) and eq.(2.58) are convenience expressions written in [2], where the partition functions in the sums are abbreviated, and their exact meaning is as follows. For example, the exact expressions of eq. (2.57) are given in eq.(2.63) below.

The partition function $Z_\delta$ here has the following form [7]:

$$Z_\delta = \frac{\Xi_6[\delta]\vartheta[\delta](0)^4}{\Psi_{10}} \tag{2.59}$$

All factors are written using $2g+1=5$ branch points $e_1, e_2, \dots e_5$ with fixing $e_6 = \infty$, by modifying the original method in [7].

$\Psi_{10}$ is equal to the square of the discriminant $\sqrt{D} = \prod_{i<j}(e_i - e_j)$ up to the overall numerical factor,

$$\Psi_{10} = (\sqrt{D})^2 = \{\prod_{i<j}(e_i - e_j)\}^2 \tag{2.60}$$

In [2], the spin sum

$$\sum_\delta \quad Z_\delta P_{IJ}(\Omega_\delta)P_{KL}(\Omega_\delta) \tag{2.61}$$

is conventionally written as

$$\frac{1}{\sqrt{D}}\sum_\delta \frac{\Xi_6[\delta]\vartheta[\delta](0)^4}{\sqrt{D}} P_{IJ}(\Omega_\delta)P_{KL}(\Omega_\delta) \tag{2.62}$$

and the sum in this $\sum_\delta$ has been carried out. The eq.(2.57) means the sums in eq.(2.62).

That is, the abbreviated expressions $\sum_\delta P_{12}P_{12} = 2\sqrt{D}$, $\sum_\delta P_{11}P_{22} = -4\sqrt{D}$ are equivalent to

$$\sum_\delta \frac{\Xi_6[\delta]\vartheta[\delta](0)^4}{\Psi_{10}} P_{12}(\Omega_\delta)P_{12}(\Omega_\delta) = 2 \quad , \quad \sum_\delta \frac{\Xi_6[\delta]\vartheta[\delta](0)^4}{\Psi_{10}} P_{11}(\Omega_\delta)P_{22}(\Omega_\delta) = -4, \tag{2.63}$$

up to the overall numerical constant. The overall numerical factor is not accurately assessed in [2].

In genus 2, one non-singular even spin structure corresponds to one choice of two branch points $e_i, e_j$. There are 10 such choices, and we denote one choice as $\delta$ or as $(i,j)$.

For a fixed value of $I,J$ and $K,L$ in $P_{IJ}(\Omega_\delta)P_{KL}(\Omega_\delta)$, choose the polynomials of branch points from the result

$P_{22}(\Omega_{ij}) = e_i + e_j, \quad P_{12}(\Omega_{ij}) = P_{21}(\Omega_{ij}) = -e_i e_j$,

$P_{11}(\Omega_{ij}) = (e_p + e_q + e_r)e_i e_j + e_p e_q e_r$

where each of $e_p,\ e_q,\ e_r$ is different from $e_i,\ e_j$ ,
and summing over

$$\frac{\Xi_6[i,j]\vartheta[i,j](0)^4}{\Psi_{10}} P_{KL}(\Omega_{i,j}) P_{IJ}(\Omega_{i,j}) \tag{2.64}$$

for all 10 choices of $(i,j)$.
In order to have non-zero results, it is necessary to have enough number of degrees as polynomials of the branch points in the numerator in (2.64). $P_{22}P_{22}$, $P_{22}P_{12}$ don't have enough degrees, and so $\sum_\delta P_{22}P_{22} = \sum_\delta P_{22}P_{12} = 0$.

As for $P_{12}P_{11}$ , the result of the sum of $\frac{\Xi_6[i,j]\vartheta[i,j](0)^4}{\Psi_{10}} P_{12}(\Omega_{i,j}) P_{11}(\Omega_{i,j})$ will be degree one symmetric polynomial of branch points. This has to have the form $e_1 + e_2 + \cdots e_5$. In genus 2, the condition $e_1 + e_2 + \cdots e_5 = 0$ will be imposed, as we did $e_1 + e_2 + e_3 = 0$ in the genus 1 case. So $\sum_\delta P_{12}P_{11} = 0$.
Therefore, only $P_{12}(\Omega_\delta)P_{12}(\Omega_\delta)$ , $P_{11}(\Omega_\delta)P_{22}(\Omega_\delta)$ and $P_{11}(\Omega_\delta)P_{11}(\Omega_\delta)$ have the possibility to give non-zero results. The results of the first two should be c number, which gives (2.63). The result of the spin sum $P_{11}(\Omega_\delta)P_{11}(\Omega_\delta)$ can give a degree two symmetric polynomial, but the result shows that it happens to give zero.

Substituting the results (2.63) into the product $[\ \sum_{I,J} P_{IJ}(\Omega_\delta)\omega_I(z_1)\omega_J(z_2)][\sum_{K,L} P_{KL}(\Omega_\delta)\omega_K(z_3)\omega_L(z_4)\ ]$, we see that the spin sum reproduces the correct combination of 1 forms

$$-2\ [\ \Delta(z_1,z_3)\Delta(z_2,z_4) + \Delta(z_1,z_4)\Delta(z_2,z_3)\ ]$$

up to overall constant, for $I_{20}(z_1,z_2,z_3,z_4)$. The spin sum calculations by multiplying partition functions in this case is, as a result, to replace each term of $P_{IJ}(\Omega_\delta)P_{KL}(\Omega_\delta)$ with 0, 2, or $-4$ in the product (2.56) ( or in any results of Wick contractions given later, say (2.66) or (2.78) ). The overall numerical constant $-2$ in the result above is inaccurately evaluated in ref.[2], compared with the accurate overall constant $-4\pi^4$ in the expression of $I_{20}(z_1,z_2,z_3,z_4)$ in (2.54) that is from ref.[7].

**Genus 2 case ,  $N = 4$ product**
In genus 2 , 4-point amplitudes, the following spin sum was also described in [7]:

$$I_{21}(z_1,z_2,z_3,z_4) = \sum_\delta \Xi_6[\delta]\vartheta[\delta](0)^4 S_\delta(z_1,z_2)S_\delta(z_2,z_3)S_\delta(z_3,z_4)S_\delta(z_4,z_1)$$

$$= +2\pi^4\Psi_{10}(\Delta(z_1,z_2)\Delta(z_3,z_4) - \Delta(z_1,z_4)\Delta(z_2,z_3)) \tag{2.65}$$

From eq. (2.48), the only terms that can be non-zero after the spin sum are

$$\sum_{I,J,K,L} P_{IJKL}(\Omega_\delta)\ \omega_I(z_1)\omega_J(z_2)\ \omega_K(z_3)\ \omega_L(z_4) \tag{2.66}$$

$P_{IJKL}$ can be expressed by $P_{MN}$ for the genus 2 as in [16]

$$P_{1111} = 6P_{11}^2 - 8\,\mu_5\mu_1 + 2\,\mu_4\mu_2 - 12\mu_5 P_{22} + 4\mu_4 P_{21} + 4\,\mu_3 P_{11}$$
$$P_{2111} = 6P_{11}P_{12} - 4\,\mu_5 - 2\,\mu_4 P_{22} + 4\mu_3 P_{21}$$
$$P_{2211} = 4P_{21}^2 + 2\,P_{22}P_{11} + 2\mu_2 P_{21}$$
$$P_{2221} = 6\,P_{22}P_{21} + 4\mu_1 P_{21} - 2P_{11}$$
$$P_{2222} = 6\,P_{22}^2 + 2\mu_2 + 4\,\mu_1 P_{22} + 4P_{21} \qquad (2.67)$$

For genus 2, the curve is expressed as

$$y^2 = \prod_{k=1}^{5}(x - e_k) = R(x) = x^5 + \mu_1 x^4 + \mu_2 x^3 + \ldots + \mu_5 \qquad (2.68)$$

Looking at equations (2.67), only the first two terms $4P_{21}^2 + 2\,P_{22}P_{11}$ on the right-hand side of $P_{2211}$ above can give the non-zero results. However, as can easily be confirmed, these terms don't give the combination

$$\Delta(z_1, z_2)\Delta(z_3, z_4) - \Delta(z_1, z_4)\Delta(z_2, z_3)\ .$$

That is, $\sum_{I,J,K,L} P_{IJKL}(\Omega_\delta)\,\omega_I(z_1)\omega_J(z_2)\,\omega_K(z_3)\,\omega_L(z_4)$ can't reproduce the desired spin sum result for $I_{21}(z_1, z_2, z_3, z_4)$.

This means that, the expansion functions

$P_{IJ}(\alpha)$, $P_{IJK}(\alpha)$, $P_{IJKL}(\alpha)$, , $P_{IJKLM}(\alpha)\ldots$, which are direct generalisations of the genus 1 case, do not have enough function space dimension, in the 4 point case in genus 2.

Denote Jacobi variety $J = \mathbb{C}^g/\Lambda$ .

Let $\Theta^{[g-1]}$ be the standard theta cycle, and $\Gamma(\mathrm{J},\ \mathcal{O}(\mathrm{n}\Theta^{[g-1]}))$ be the function space on $J$ formed by the whole of the functions which have at most order $n$ poles[6] in $\Theta^{[g-1]}$ and are regular at all other points.

Then, the following dimension theorem holds as in ref [16]:

$$\text{For } n \geq 2, \quad \dim \quad \Gamma\left(\mathrm{J}, \mathcal{O}\left(\mathrm{n}\Theta^{[g-1]}\right)\right) = n^g\ . \qquad (2.69)$$

For the genus 2 for example, the following formulae are also known:

$$\Gamma\left(\mathrm{J}, \mathcal{O}\left(2\,\Theta^{[1]}\right)\right) = \mathbb{C}\,1 \oplus \mathbb{C}P_{11} \oplus \mathbb{C}P_{12} \oplus \mathbb{C}P_{22} \qquad \left(\,dim\Gamma\left(\mathrm{J}, \mathcal{O}\left(2\,\Theta^{[1]}\right)\right) = 2^2 = 4\,\right) \qquad (2.70)$$

$$\Gamma\left(\mathrm{J}, \mathcal{O}\left(3\,\Theta^{[1]}\right)\right) = \Gamma\left(\mathrm{J}, \mathcal{O}\left(2\,\Theta^{[1]}\right)\right) \oplus \mathbb{C}\,P_{111} \oplus \mathbb{C}P_{112} \oplus \quad \mathbb{C}P_{122} \oplus \quad \mathbb{C}P_{222} \oplus \mathbb{C}(P_{12}^2 - P_{11}P_{22})$$

$$\left(\,dim\Gamma\left(\mathrm{J}, \mathcal{O}\left(3\,\Theta^{[1]}\right)\right) = 3^2 = 9\,\right) \qquad (2.71)$$

---

[6] To say that a function F has an order n pole here means that F has an order n pole in the usual sense on the section (which is one dimension in $\mathbb{C}$ ) of a surface in $\mathbb{C}^g/\Lambda$ .

In higher genus the number of base functions to expand the product $\Pi_{i=1}^{N}\ F_g(A(z_i - z_{i+1}),\alpha)$ increases rapidly as $N$ becomes large. Here we prepare all such base functions with appropriate order of poles, such as

$$P_{AB}(\alpha),\ P_{ABC}(\alpha),\ P_{AB}(\alpha)P_{CD}(\alpha),\ P_{ABC}(\alpha)P_{DE}(\alpha),\ P_{AB}(\alpha)\,P_{CD}(\alpha)P_{EF}(\alpha)\ldots\ldots \qquad (2.72)$$

For the even order of poles, we prepare the product of $P_{AB}(\alpha)$ s. For the odd order of poles, we multiply the odd order pole functions, $P_{ABC}(\alpha), P_{ABCDE}(\alpha)$ , ... to the product of $P_{AB}(\alpha)$ s.
We assume that these base functions give enough functional space to expand the function $\Pi_{i=1}^{N}\ F_g(A(z_i - z_{i+1}),\alpha)$ under the cyclic condition $z_{i+1} = z_1$ . In genus 1, there is a different way of expansion from that of (1.24), and it has similar features of these expansion base functions. We have briefly described these in appendix B.

Up to $N = 3$, only $P_{AB}(\alpha)$ , $P_{ABC}(\alpha)$ are needed as such base functions, so the form of the expansion formula will be the same as in the last chapter:

$$S_\delta(z_1,z_2)S_\delta(z_2,z_1) = -S_\delta(z_1,z_2)^2 = P_{IJ}(A(z_1 - z_2))\omega_I(z_1)\omega_J(z_2) + P_{IJ}(\Omega_\delta)\omega_I(z_1)\omega_J(z_2) \qquad (2.73)$$

$$\begin{aligned} &S_\delta(z_1,z_2)S_\delta(z_2,z_3)S_\delta(z_3,z_1) \\ &= \{\, \partial_{IJK}\, L_3 \}\, \omega_I(z_1)\omega_J(z_2)\omega_K(z_3) + \{\ [\,\partial_I\, L_3][\, D_{JK}^2\, L_3] + [\,\partial_J\, L_3][\, D_{IK}^2\, L_3] + [\,\partial_K\, L_3][\, D_{IJ}^2\, L_3] \\ &\qquad + [\,\partial_I\, L_3][\,\partial_J\, L_3][\,\partial_K\, L_3]\, \}\, \omega_I(z_1)\omega_J(z_2)\omega_K(z_3) \\ &\quad + \{\ [\,\partial_I\, L_3][\, P_{JK}(\Omega_\delta)] + [\,\partial_J\, L_3][P_{IK}(\Omega_\delta)\,] + [\,\partial_K\, L_3][\, P_{IJ}(\Omega_\delta)]\ \}\, \omega_I(z_1)\omega_J(z_2)\omega_K(z_3) \end{aligned} \qquad (2.74)$$

where the numerical factors in each term are all abbreviated because it is difficult to determine them without clarifying the pole structures of $ln\theta_R(u)$ .

As for $N = 4,5$, we need $P_{AB}(\alpha)$ , $P_{ABC}(\alpha)$, $P_{AB}(\alpha)P_{CD}(\alpha)$ as the base functions of the expansion, in which $A,B,C,D$ will be contracted with $I,J,K,L$ , and the result may be written as follows after setting $\alpha = \Omega_\delta$ :

$$\begin{aligned} &S_\delta(z_1,z_2)S_\delta(z_2,z_3)S_\delta(z_3,z_4)S_\delta(z_4,z_1) \\ &\quad = \tilde{\tilde{H}}_{IJKM}\ \ \omega_I(z_1)\omega_J(z_2)\omega_K(z_3)\omega_M(z_4) \\ &+\{\, \{[\partial_I L_4]\,[\partial_J L_4] + [D_{IJ}^2\, L_4]\}[P_{KM}(\Omega_\delta)] + (cycl\ I,J,K,M)\, \}\, \omega_I(z_1)\omega_J(z_2)\omega_K(z_3)\omega_M(z_4) \\ &+\ P_{AB}(\Omega_\delta)P_{CD}(\Omega_\delta)\omega_I(z_1)\omega_J(z_2)\omega_K(z_3)\omega_M(z_4) \end{aligned} \qquad (2.75)$$

The details of the spin structure independent term $\tilde{\tilde{H}}_{IJKM}$ are not clear at present.

For N=5, the result may be written as follows:

$$\begin{aligned} &S_\delta(z_1,z_2)S_\delta(z_2,z_3)S_\delta(z_3,z_4)S_\delta(z_4,z_5)S_\delta(z_5,z_1) \\ &= \tilde{\tilde{H}}_{IJKMN}\ \ \omega_I(z_1)\omega_J(z_2)\omega_K(z_3)\omega_M(z_4)\omega_N(z_5) \end{aligned}$$

$$+\{ X_{IJK}P_{MN}(\Omega_\delta) + (cycl\ I,J,K,M,N) \}\ \omega_I(z_1)\omega_J(z_2)\omega_K(z_3)\omega_M(z_4)\omega_N(z_5)$$
$$+ \{ \partial_I L_5\ P_{AB}(\Omega_\delta)P_{CD}(\Omega_\delta) + (cycl\ I,A,B,C,D) \}\omega_I(z_1)\omega_J(z_2)\omega_K(z_3)\omega_M(z_4)\omega_N(z_5) \quad (2.76)$$

where

$$X_{IJK} = \partial_{IJK}L_5 + [\partial_I L_5][D^2_{JK}L_5] + [\partial_J L_5][D^2_{IK}L_5] + [\partial_K L_5][D^2_{IJ}L_5] + [\partial_I L_5][\partial_J L_5][\partial_K L_5] \quad (2.77)$$

which may have the same form as the one constructed using base expansion functions in the last chapter.

The terms which can give non-zero contributions by the spin sum of $I_{21}(z_1,z_2,z_3,z_4)$ in 4-point amplitude of genus 2 in (2.75) are, instead of (2.66),

$$\sum \quad P_{AB}(\Omega_\delta)P_{CD}(\Omega_\delta)\ \omega_I(z_1)\omega_J(z_2)\ \omega_K(z_3)\ \omega_M(z_4) \quad (2.78)$$

where the indices $A,B,C,D$ are to be contracted with $I,J,K,M$ .

The only one possible way of contraction that can give accurate results which is also consistent with $I_{11}(z_i) = I_{12}(z_i)$ in eq.(3.3) in ref.[7] is to write $A,B,C,D$ as[7] :

$$\sum_{I,J,K,L} \frac{1}{2}\{P_{IJ}(\Omega_\delta)P_{KL}(\Omega_\delta) + P_{LI}(\Omega_\delta)P_{JK}(\Omega_\delta)\}\ \omega_I(z_1)\omega_J(z_2)\ \omega_K(z_3)\ \omega_L(z_4) \quad (2.79)$$

Picking up non-zero contributions to the spin sum reads

$$\begin{aligned}
&+\textstyle\sum_\delta \frac{1}{2}(P_{11}P_{22} + P_{12}P_{12})\ \ \omega_1(z_1)\omega_1(z_2)\ \omega_2(z_3)\ \omega_2(z_4)\\
&+\textstyle\sum_\delta \frac{1}{2}(P_{21}P_{12} + P_{22}P_{11})\ \omega_2(z_1)\omega_1(z_2)\ \omega_1(z_3)\ \omega_2(z_4)\\
&+\textstyle\sum_\delta \frac{1}{2}(P_{12}P_{21} + P_{11}P_{22})\ \omega_1(z_1)\omega_2(z_2)\ \omega_2(z_3)\ \omega_1(z_4)\\
&+\textstyle\sum_\delta \frac{1}{2}(P_{22}P_{11} + P_{21}P_{21})\ \omega_2(z_1)\omega_2(z_2)\ \omega_1(z_3)\ \omega_1(z_4)\\
&+\textstyle\sum_\delta \frac{1}{2}(P_{12}P_{12} + P_{21}P_{21}\ )\ \omega_1(z_1)\omega_2(z_2)\ \omega_1(z_3)\ \omega_2(z_4)\\
&+\textstyle\sum_\delta \frac{1}{2}(P_{21}P_{21} + P_{12}P_{12})\ \omega_2(z_1)\omega_1(z_2)\ \omega_2(z_3)\ \omega_1(z_4)
\end{aligned} \quad (2.80)$$

Replacing $\sum_\delta P_{12}P_{12}$ with 2, $\sum_\delta P_{11}P_{22}$ with $-4$ in (2.80) gives the correct combination

$$+1\ [\ \Delta(z_1,z_2)\Delta(z_3,z_4) - \Delta(z_1,z_4)\Delta(z_2,z_3)]\ , \quad (2.81)$$

using the symmetry $P_{KL} = P_{LK}$ .

Here let us look at genus 2 case (2.67) again:

$$P_{1111} = 6P_{11}^2 - 8\,\mu_5\mu_1 + 2\,\mu_4\mu_2 - 12\mu_5P_{22} + 4\mu_4P_{21} + 4\,\mu_3P_{11}$$

---

[7] The author thanks to the referee for not only pointing out the errors in the descriptions of this point in the previous versions of the manuscript, but also for explaining the way to the solution.

$P_{2111} = 6P_{11}P_{12} - 4\,\mu_5 - 2\,\mu_4 P_{22} + 4\mu_3 P_{21}$

$P_{2211} = 4P_{21}^2 + 2\,P_{22}P_{11} + 2\mu_2 P_{21}$

$P_{2221} = 6\,P_{22}P_{21} + 4\mu_1 P_{21} - 2P_{11}$

$P_{2222} = 6\,P_{22}^2 + 2\mu_2 + 4\,\mu_1 P_{22} + 4P_{21}$

In genus 2, there are five $P_{IJKL}$ and six $P_{MN}P_{QR}$ and, as in (C), $P_{IJKL}$ can be represented by $P_{MN}P_{QR}$ and lower degree $P_{IJ}$ terms. Looking the other way around, $P_{11}^2$, $P_{22}^2$, $P_{22}P_{21}$, $P_{11}P_{12}$ can be represented by $P_{IJKL}$. Only $P_{22}P_{11}$ *and* $P_{21}^2$, which happen to give non-zero spin sums, are combined in $P_{2211}$. This shows an example of the difference in the dimension of the space of expansion base functions.

In the cases $N = 4$ and 5, in the spin structure independent terms $\overset{\frown}{H}_{IJKM}$ and $\overset{\frown}{H}_{IJKMN}$ in (2.75) and (2.76), zero modes of the base functions of the expansions $P_{IJ}(\alpha)$, $P_{IJK}(\alpha)$, $P_{IJ}(\alpha)P_{KL}(\alpha)$ will be included, as described in subsection 2.3.

.

**Higher points**

For $N = 5$, in the case of genus 2, the following spin sums are calculated in [8]:

$$J_1(z_1, z_2, z_3, z_4, z_5) = \sum_{\delta} \Xi_6[\delta]\vartheta[\delta](0)^4 S_\delta(z_1, z_2)S_\delta(z_2, z_3)S_\delta(z_3, z_1)S_\delta(z_4, z_5)^2$$

$$= -2\pi^4\Psi_{10}\omega_I(z_1)\big(\Delta(z_2, z_4)\Delta(z_3, z_5)G^I_{1,3,5,4,2} + (4 \leftrightarrow 5)\big) + \mathrm{cycl}(1,2,3) \qquad (2.82)$$

$$J_2(z_1, z_2, z_3, z_4, z_5) = \sum_{\delta} \Xi_6[\delta]\vartheta[\delta](0)^4 S_\delta(z_1, z_2)S_\delta(z_2, z_3)S_\delta(z_3, z_4)S_\delta(z_4, z_5)\,S_\delta(z_5, z_1)$$

$$= -2\pi^4\Psi_{10}\omega_I(z_1)\Delta(z_2, z_5)\Delta(z_3, z_4)G^I_{5,1,2} + \mathrm{cycl}(1,2,3,4,5) \qquad (2.83)$$

$J_1$ corresponds to the case $(N = 3) \times (N = 2)$. Only the following terms will give non-zero contributions after the spin sum:

$$[\,\textstyle\sum_{IJK} \{[\partial_I\, L_3]\big[\,P_{JK}(\Omega_\delta)\big] + [\,\partial_J\, L_3][P_{IK}(\Omega_\delta)\,] + [\,\partial_K\, L_3]\big[\,P_{IJ}(\Omega_\delta)\big]\ \}\,\omega_I(z_1)\omega_J(z_2)\omega_K(z_3)\,] \times$$
$$[\,\textstyle\sum_{K,L} P_{KL}(\Omega_\delta)\omega_K(z_4)\omega_L(z_5)] \qquad (2.84)$$

$J_2$ corresponds to the $N = 5$ case. Only the following terms will give non-zero contributions

$$\sum_{I,J,K,L,M} [\,\partial_I L_5]\,\frac{1}{2}\{P_{JK}(\Omega_\delta)P_{LM}(\Omega_\delta) + P_{MJ}(\Omega_\delta)P_{KL}(\Omega_\delta)\}\,\omega_I(z_1)\omega_J(z_2)\,\omega_K(z_3)\,\omega_L(z_4)\,\omega_M(z_5)$$
$$+(cycl\ I, J, K, L, M) \qquad (2.85)$$

The term $\partial_I L_5$ in (2.85) corresponds to the function $V_1(x_1, x_2, \ldots . x_5)$ in the case of genus 1.

As for the spin sum, the same patterns of a product of two $P_{IJ}(\Omega_\delta)$ appeared as those in the case of 4-point amplitude.

The combination of the holomorphic 1 form reproduces the following Eq. (3.7) in ref. [8]:

$$J_1 = -2\pi^4\Psi_{10}\frac{x_{12}dx_3}{x_{23}x_{31}}\ \left(\Delta(1,4)\Delta(2,5) + (4 \leftrightarrow 5)\right) + \mathrm{cycl}(1,2,3) \tag{2.86}$$

$$J_2 = -\pi^4\Psi_{10}\frac{x_{14}dx_5}{x_{45}x_{51}}\ \left(\Delta(1,2)\Delta(3,4) - \Delta(1,4)\Delta(2,3)\right) + \mathrm{cycl}(1,2,3,4,5) \tag{2.87}$$

N=6

The result will be

$$\begin{aligned}
&S_\delta(z_1,z_2)S_\delta(z_2,z_3)S_\delta(z_3,z_4)S_\delta(z_4,z_5)S_\delta(z_5,z_6)S_\delta(z_6,z_1) = \prod_{i=1}^{6} S_\delta(x_i)\\
&= \tilde{\tilde{H}}_{IJKMNQ}\ \ \omega_I(z_1)\omega_J(z_2)\omega_K(z_3)\omega_M(z_4)\omega_N(z_5)\omega_Q(z_6)\\
&+\{\, X_{IJKM}P_{NQ}(\Omega_\delta) + (cycl\,)\,\}\ \omega_I(z_1)\omega_J(z_2)\omega_K(z_3)\omega_M(z_4)\omega_N(z_5)\omega_Q(z_6)\\
&+\{\, X_{IJ}\frac{1}{2}[P_{KM}(\Omega_\delta)P_{NQ}(\Omega_\delta)+P_{QK}(\Omega_\delta)P_{MN}(\Omega_\delta)] +\\
&\quad (cycl\,)\,\}\omega_I(z_1)\omega_J(z_2)\omega_K(z_3)\omega_M(z_4)\omega_N(z_5)\omega_Q(z_6)\\
&+ P_{AB}(\Omega_\delta)P_{CD}(\Omega_\delta)P_{EF}(\Omega_\delta)\omega_I(z_1)\omega_J(z_2)\omega_K(z_3)\omega_M(z_4)\omega_N(z_5)\omega_Q(z_6)
\end{aligned} \tag{2.88}$$

$X_{IJ}$ will be

$$X_{IJ} = [\,\partial_I L_6][\,\partial_J L_6] + \left[\,D_{IJ}^2\, L_6\right] \tag{2.89}$$

The indices $A,B,C,D,E,F$ in the Pe functions is to be contracted with $I,J,K,M,N,Q$ .

In the case that the partition function is of the type (2.59), after the spin sum, the last two lines of (2.88), the terms containing $\frac{1}{2}\{P_{KM}(\Omega_\delta)P_{NQ}(\Omega_\delta)+P_{QK}(\Omega_\delta)P_{MN}(\Omega_\delta)\}$ and $P_{AB}(\Omega_\delta)P_{CD}(\Omega_\delta)P_{EF}(\Omega_\delta)$ can give non-zero results. The results of the former may include the Pe function in the term $D_{IJ}^2\, L_6$ in (2.89), as in the case of genus one, if it is allowed to set the odd theta functions in the prime form to $ln\theta_R((2M\ \ )^{-1}u)$. In genus one, the term including $P_{AB}(\Omega_\delta)P_{CD}(\Omega_\delta)P_{EF}(\Omega_\delta)$ was zero after the spin sum.

Unfortunately, we can't give logically definite way of how in general the indices of the Pe functions factors $P_{AB}(\Omega_\delta)P_{CD}(\Omega_\delta)P_{EF}(\Omega_\delta)$ should be contracted in this paper[8].

Also, we can't guess the form of $X_{IJKM}$ . At higher point amplitudes, because the pole structures of $P_{IJ}(\alpha)P_{KM}(\alpha)P_{NQ}(\alpha)\ldots$ becomes complicated, it becomes difficult at present to guess the form of the expansion coefficients. (See Appendix B.) The $X_{IJKM}$ may include "the second zero mode" $\Lambda_{IJKM}$ , and may include combinations of terms that appeared in the $N = 4$ case.

In general, to clarify the expansion coefficients of the product $\Pi_{i=1}^{N} F_g(u_i,\alpha)$

[8] After submitting this manuscript to ArXiv, I found a very impressive paper ref.[20], in which general method of summing over spin structures at genus 2 is described in detail.

defined in (2.20), we first need to know the mathematical structure of the expansion form of $ln\theta_R(u)$ .

As for the method of summing over spin structures, it seems we can find an effective way for at least for genus 2 and hyper-elliptic genus 3 cases.

For these cases, as described at the end of Appendix B, $P_{I_1 I_2 \ldots I_K}$ for any positive integer K>1 can be written only by $P_{I_1 I_2}$ and $P_{I_1 I_2 I_3}$ . All the spin structure dependent terms in $\Pi_{i=1}^N S_\delta(z_i,\ z_{i+1})$ will have the form of the direct products (simple products) of $P_{IJ}(\Omega_\delta)$, that is, $P_{IJ}(\Omega_\delta)$, $P_{IJ}(\Omega_\delta)P_{KL}(\Omega_\delta)$, $P_{IJ}(\Omega_\delta)P_{KL}(\Omega_\delta)P_{MN}(\Omega_\delta)$,.. .
For $N = 2, 3$ , the product $\Pi_{i=1}^N S_\delta(z_i,\ z_{i+1})$ has spin structure dependent terms proportional to $P_{IJ}(\Omega_\delta)$ . For $N = 4, 5$ , $P_{IJ}(\Omega_\delta)$ and $P_{IJ}(\Omega_\delta)P_{KL}(\Omega_\delta)$ ; for $N = 6,7$ , $P_{IJ}(\Omega_\delta)$ and $P_{IJ}(\Omega_\delta)P_{KL}(\Omega_\delta)$ and $P_{IJ}(\Omega_\delta)P_{KL}(\Omega_\delta)P_{MN}(\Omega_\delta)$, and so on.
Then, by the fact (C), the spin structure dependent terms can only be expressed by branch points. We note again that the (C) is valid only in the hyper-elliptic cases.

In general cases beyond genus 2 and hyperelliptic genus 3 curves, it is still an open problem whether the base functions of the expansion can be restricted to the products of $P_{IJ}$ and $P_{KLI}$ as in Appendix B. If higher rank tensors $P_{IJKLM\ldots}$ , which are not decomposable into $P_{IJ}$ and $P_{KLI}$ , are to be included in the base functions, the theorem in (C) cannot be used.

In the cases that the (C) can be applied, in the language of theta constants, the spin structure dependence of the product $\Pi_{i=1}^N S_\delta(z_i,\ z_{i+1})$ with cyclic condition has the form of the simple products of
$-[2\eta(2M)^{-1}]_{IJ} - [\,((2M)^{-1})^T H (2M)^{-1}]_{IJ}$ , by the theorem (2.26).
When the decomposition formula is applied to string amplitudes, partition functions $Z_\delta$, which will also be functions of branch points, are multiplied by the simple products of $P_{IJ}(\Omega_\delta)$ and summed over $\delta$ . This spin sum is a simple algebra of only $2g + 1$ branch points, and there is no difficulty in principle for arbitrary $N$ by the procedure already described, although the computations will be quite long for large values of $N$ or $g$ ( in hyperelliptic cases) .
The general results of the spin sum will be symmetric functions of the branch points $e_1$ , $e_2, \ldots e_{2g+1}$, as in the case of genus 1. These symmetric functions will be modular covariant combinations of generalised holomorphic Eisenstein series. All zero modes in the expression of the coefficients written in terms of $W_{I_1, I_2, \ldots I_K}(z_1, z_2, \ldots . z_N)$ will also be generalised holomorphic Eisenstein series.

## 3. Summary and Conclusions

In this paper, we have tried to clarify some properties of the simple product of fermion correlation functions

$S_\delta(z_1, z_2)S_\delta(z_2, z_3).....S_\delta(z_N, z_{N+1})$ with the constraint $z_{N+1} = z_1$. (1.2)

to apply to simplified RNS formalism calculations of superstring amplitudes.

As is well known, for higher genus $g \geq 2$ , as clarified in [7] and related papers, even in the case of calculating amplitudes of external massless bosons in Type I and Type II superstrings, we need to include other types of vertex operators in which the cyclic conditions are not satisfied. The main purpose of this paper was to consider an efficient way of summing over spin structures of the type with cyclic condition only.

In genus 1, under such cyclic condition, the spin structure dependent part of the product can be extracted as $P^{(2K-2)}(\omega_\delta)$ , which does not depend on the vertex insertion points $z_1, z_2, \dots z_N$ , and we can give an efficient method of summing over spin structures. Including all the cases as described in chapter1, this can be summarised in a compact closed form (1.11) or (2.52) as

$$\prod_{i=1}^{N} S_\delta(x_i) = \sum_{K=0}^{\left[\frac{N}{2}\right]} H_{N,N-2K}\, P^{(2K-2)}(\omega_\delta)$$

$$= V_N - \sum_{K=1}^{\left[\frac{N}{2}\right]} \frac{1}{(2K-1)!} V_{N-2K} \cdot D^{2K} \ln \theta_1(\omega_\delta)$$

where $H_{N,M}$ are given by in (1.12) and (1.13), $D^{2K} \ln \theta_1(\omega_\delta)$ is given in (2.53).

When this formula is used to calculate the superstring amplitudes, the partition functions $Z_\delta$ are multiplied by $P^{(2K-2)}(\omega_\delta)$ . The results of the spin sum $\sum_\delta Z_\delta P^{(2K-2)}(\omega_\delta)$ are generally expressed by holomorphic Eisenstein series after algebraic calculations of the branch points.

For hyperelliptic cases, as for non-singular even spin structures, we derived a proposition in (2.15) which is valid in any genus. A key theorem (2.26) is also for non-singular and even spin structures.

If we define a function $F_g(u, \alpha) \stackrel{\text{def}}{=} \frac{\theta_R(u+\alpha)}{\theta_R(\alpha)E(z,w)}$ where $\theta_R(u)$ is a theta function of genus $g$ whose characteristics are components of the vector of Riemann constants, we have

$$\Pi_{i=1}^{N} S_\delta(z_i,\ z_{i+1}) = \Pi_{i=1}^{N} F_g(A(z_i - z_{i+1}),\ \Omega_\delta) = \Pi_{i=1}^{N} \frac{\theta_R(A(z_i - z_{i+1}) + \Omega_\delta)}{\theta_R(\ \Omega_\delta)E(z_i, z_{i+1})} .$$

In this sense, $F_g(u, \alpha)$ can be regarded as a generalization of the Eisenstein- Kronecker series in higher genus for our purposes.

In genus 1, the Pe function is expressed by $\theta_1(x)$ and $\eta_1$, and the product of the fermion correlation functions with cyclic condition is expressed by $\theta_1(x)$ and $\omega_\delta$ through Eisenstein-Kronecker series, as in (1.23). As for the higher genus cases we are concerned with, both can be expressed by replacing $\theta_1(x)$ by $\theta_R(u)$, $\eta_1$ by $\eta_{IJ}$, $\omega_\delta$ by $\Omega_\delta$, by the definition of the Pe function and (2.15). This generalisation has such a simple form as a result. In genus 1, by expanding the product (1.22), $\Pi_{i=1}^{N}\frac{\theta_1^{(1)}(0)\theta_1(x_i+\alpha,\tau)}{\theta_1(x_i,\tau)\theta_1(\alpha,\tau)}$ by Pe functions, we were able to have simple closed form of the decomposition formula of $\Pi_{i=1}^{N}S_\delta(x_i)$ for arbitrary $N$. Similarly, rigorous arguments for the expansion of the product (2.20), $\Pi_{i=1}^{N}\frac{\theta_R(A(z_i-z_{i+1})+\alpha)}{\theta_R(\alpha)E(z_i,z_{i+1})}$ by the Pe functions $P_{IJ}(\alpha)$ may also give an efficient way to derive the decomposition formula of the fermion correlation functions in some cases of higher genus and the spin sum method for arbitrary $N$. The author hopes that a formula similar to (2.15) will hold for general Riemann surfaces.

The facts and assumptions as (B) ~ (E) described below eq.(2.19) are expected to hold mainly for the hyperelliptic cases. Therefore, as described there, only the case of genus 2 is realistic when the method discussed in chapter 2 is applied to the string theories at present. In particular, except genus 2 and hyperelliptic genus 3 cases, it is not yet certain whether $P_{I_1 I_2 \dots I_K}$ for positive integer K>3 can be written only by $P_{I_1 I_2}$ and $P_{I_1 I_2 I_3}$ or not.

In the method described in chapter 2, a point is that, by setting $\alpha = \Omega_\delta$, the functions of $P_{IJ}(\alpha)$ in the expansion base functions become the spin structure dependent parts of the product of $\Pi_{i=1}^{N}S_\delta(z_i, z_{i+1})$, which are independent of the vertex insertion points, while the coefficients of the expansion become modular covariant functions of the vertex insertion points which remain unaffected by the spin sum process.

The main problem in pursuing this at present is that we don't know the structure of the Laurent expansion of $ln\theta_R(u)$. Obviously, we need to know the higher genus generalisation of the following function formula of Pe :

$$P(\alpha) \equiv -\frac{d^2 \ln \sigma}{d\alpha^2} = \alpha^{-2} + \sum_{m=1}^{\infty}(2m+1)G_{2m+2}(\tau)\alpha^{2m} \quad . \qquad \text{(A.8)}$$

At present we don't have a logical method to obtain the exact forms of the coefficients of the expansion of $\Pi_{i=1}^{N}\ F_g(A(z_i - z_{i+1}), \alpha)$ for genus $g \geq 2$. In Chapter 2, based on the analogy of the case of genus 1, we discussed the possible function forms, by paying attention to the way how the factor $ln\theta_R(\alpha)$ in the denominator of the product $\Pi_{i=1}^{N}\ F_g(A(z_i - z_{i+1}), \alpha)$ was treated. We had to restrict our arguments to the case where $\theta_R(\alpha)$ is odd, and set the odd theta function in the prime form to the $\theta_R(\alpha)$. Although there is a possibility that in the rigorous treatment of obtaining the coefficient functions of the expansion of $\Pi_{i=1}^{N} F_g(A(z_i - z_{i+1}), \alpha)$ the concepts described in Chapter 2 may be replaced by more sophisticated ones, we expect that the descriptions given there will give some hints for future investigations.

In particular, the following seems plausible, at least for hyper-elliptic case :

If we allow the expansion of $\Pi_{i=1}^{N} F_g(A(z_i - z_{i+1}), \alpha)$ by combinations of higher genus Pe functions as base functions, the spin structure dependent terms of the decomposition formula of $S_\delta(z_1, z_2)S_\delta(z_2, z_3)\ldots\ldots S_\delta(z_N, z_{N+1})$ with cyclic condition will have the form of direct products of $P_{IJ}(\Omega_\delta)$, that is,

$$P_{IJ}(\Omega_\delta),\quad P_{IJ}(\Omega_\delta)P_{KL}(\Omega_\delta),\quad P_{IJ}(\Omega_\delta)P_{KL}(\Omega_\delta)P_{MN}(\Omega_\delta), \ldots$$

Then the method described in chapter 2 will give a concrete way to compute the summation over the non-singular even spin structure for arbitrary $N$ naturally, only through elementary algebra of fundamental symmetric functions of the branch points. The calculations of summing over spin structures will not present any substantial difficulties, whatever their concrete forms may be. This calculation will also give different combinations of holomorphic 1 forms in the results. The result will be symmetric functions of the branch points. In this paper, we temporary called all of coefficients of Laurant expansion of $ln\theta_R(u)$ as "generalised Eisenstein series", and in the result of the spin sum any symmetric functions of branch points will be expressed as modular covariant combinations of those, which can also be expressed by theta constants.

## Appendix A Notations of genus one

Half periods $\omega_\delta$ are related to moduli parameter as :

$$\omega_1 = \frac{1}{2},\quad \omega_2 = -\frac{1+\tau}{2},\quad \omega_3 = \frac{\tau}{2} \tag{A.1}$$

Periods: $A_{m,n} = 2m\omega_1 + 2n\omega_3$ (A.2)

where m, n are integers.

Starting from the infinite product representation of sigma function at genus 1,

$$\sigma(\alpha) = \alpha \cdot \prod_{m,n}' \left(1 - \frac{\alpha}{A_{m,n}}\right) exp\ \left[\frac{\alpha}{A_{m,n}} + \frac{\alpha^2}{2{A_{m,n}}^2}\right] \tag{A.3}$$

taking log of sigma function and expanding the terms which do not have singular part at $\alpha = 0$ , we have

$$\ln\sigma(\alpha) = \ln\alpha - \sum_{k=2}^{\infty} \frac{G_{2k}}{2k} \alpha^{2k} \tag{A.4}$$

where $G_{2k}$ are holomorphic Eisenstein series defined in (      ), which can also be written

as

$$G_{2k}(\tau) \equiv \sum_{(m,n)\neq(0,0)} A_{m,n}^{-2k} \tag{A.5}$$

Then unique odd theta function on the torus can be written as, adding $G_2$ ,

$$\ln\theta_1(\alpha) = \ln\sigma(\alpha) - \frac{G_2}{2}\alpha^2 + \ln\theta_1^{(1)}(0) = \ln\alpha - \sum_{k=1}^{\infty}\frac{G_{2k}}{2k}\alpha^{2k} + \ln\theta_1^{(1)}(0) \tag{A.6}$$

Also, we define

$$\varsigma(z) = \frac{d\ln\sigma(z)}{dz} \tag{A.7}$$

$$P(\alpha) \equiv -\frac{d^2\ln\sigma}{d\alpha^2} = -2\eta_1 - \frac{\partial^2}{\partial\alpha^2}\ln\theta_1(\alpha,\tau) = \alpha^{-2} + \sum_{m=1}^{\infty}(2m+1)G_{2m+2}(\tau)\alpha^{2m} \tag{A.8}$$

$$P^{(2n)}(\alpha) = \frac{(2n+1)!}{\alpha^{2n+2}} + (2n+1)!\,G_{2n+2}(\tau) + O(\alpha)\ldots\ldots \tag{A.9}$$

The values of Pe function at half periods are the branch points of the curve:

$$e_\delta = P(\omega_\delta) \qquad (\delta = 1,2,3) \tag{A.10}$$

The branch points $e_\delta$ are related to theta constants as

$$e_1 = \frac{\pi^2(\theta_4^4(0)+\theta_3^4(0))}{3}\; e_2 = \frac{\pi^2(\theta_2^4(0)-\theta_4^4(0))}{3}\;\; e_3 = \frac{-\pi^2(\theta_3^4(0)+\theta_2^4(0))}{3} \tag{A.11}$$

and

$$e_1 + e_2 + e_3 = 0 \quad e_1e_2 + e_2e_3 + e_3e_1 = -\frac{g_2}{4} \quad e_1e_2e_3 = \frac{g_3}{4} \tag{A.12}$$

Also, we use

$$\eta_\delta = \varsigma(\omega_\delta) \quad (\delta = 1,2,3). \tag{A.13}$$

$\eta_1$ is related to appropriately regularized form of $G_2(\tau)$ as

$$G_2 = 2\eta_1 \tag{A.14}$$

The $g_2, g_3$ are classical notations of modular forms which are related to holomorphic Eisenstein series $G_{2k}(\tau)$ as

$$g_2 = 60\sum_{m,n} A_{m,n}^{-4} \qquad g_3 = 140\sum_{m,n} A_{m,n}^{-6} \tag{A.15}$$

$$P^{(2n)}(z) = \frac{d^{2n}P(z)}{dz^{2n}} = polynomial\ of\ P(z)\ of\ degree\ \ n+1 \tag{A.16}$$

$$P^{(2n+1)}(z) = P^{(1)}(z) * [polynomial\ of\ P(z)\ of degree\ n\ ] \tag{A.17}$$

$$P^{(1)}(\omega_\delta) = 0 = P^{(ODD)}(\omega_\delta) \tag{A.18}$$

A few examples of (A.16) are:

$P^{(2)} = 6P^2 - \frac{1}{2}g_2$

$P^{(4)} = 120P^3 - 18g_2P - 12g_3$

This series of polynomials is consecutively constructed by differentiating

$\{P^{(1)}(z)\}^2 = 4\{P(z)\}^3 - g_2P(z) - g_3$

The coefficient of the highest degree term in $P^{(2n-2)}$ is $(2n-1)!$ .

$$P^{(2n-2)} = (2n-1)!\,P^{(n)} + \cdots$$

In the following we use matrices and determinants in which the same function is lined up in each column, and the same variable of the functions are lined up in each row.

$$\begin{vmatrix} f_1(x_1) & f_2(x_1) & \cdots & f_N(x_1) \\ f_1(x_2) & f_2(x_2) & \cdots & f_N(x_2) \\ \vdots & \vdots & & \vdots \\ \vdots & \vdots & & \vdots \\ f_1(x_N) & f_2(x_N) & \cdots & f_N(x_N) \end{vmatrix} \tag{A.19}$$

This determinant will be denoted by

$$det_{NxN}[f_1, f_2, \ldots\ldots f_N]\,(x_1, x_2, \ldots . x_N). \tag{A.20}$$

We also use the following notation, where all elements of the M-th column from the left are replaced with 1,

$$det_{NxN}[f_1, f_2, \ldots\ldots f_N]\,(M\,t\hbar \to 1;\; x_1, x_2, \ldots . x_N)$$

$$\stackrel{\text{def}}{=} \begin{vmatrix} f_1(x_1) & \cdots & f_{M-1}(x_1) & 1 & f_{M+1}(x_1) & \cdots & f_N(x_1) \\ f_1(x_2) & \cdots & f_{M-1}(x_2) & 1 & f_{M+1}(x_2) & \cdots & f_N(x_2) \\ \vdots & & & & & & \vdots \\ \vdots & & & & & & \vdots \\ f_1(x_N) & \cdots & f_{M-1}(x_N) & 1 & f_{M+1}(x_N) & \cdots & f_N(x_N) \end{vmatrix} \tag{A.21}$$

An N-th derivative of a function is always denoted as $f^{(N)}(x)$ .

Using these notations, $H_{N,N}$ and $H_{N,N-2K}$ $(K > 0)$ in (1.11) allows determinant formula as below.

$$H_{N,N}$$

$$= \frac{1}{(N-1)!} \frac{det_{(N-1)\times(N-1)} \quad [P,\; P^{(1)}, P^{(2)}, P^{(3)}, \ldots\ldots \quad P^{(N-2)}](\; x_1, x_2, \ldots x_{N-1})}{det_{(N-1)\times(N-1)} \quad [P,\; P^{(1)}, P^{(2)}, P^{(3)}, \ldots\ldots \quad P^{(N-2)}]((N-1)t\hbar \to 1;\; x_1, x_2, \ldots x_{N-1})} \tag{A.22}$$

$$H_{N,N-2K}$$

$$= \frac{-1}{(N-1)!} \frac{det_{(N-1)\times(N-1)} \quad [P,\; P^{(1)}, P^{(2)}, P^{(3)}, \ldots\ldots \quad P^{(N-2)}]((2K-1)t\hbar \to 1;\; x_1, x_2, \ldots x_{N-1})}{det_{(N-1)\times(N-1)} \quad [P,\; P^{(1)}, P^{(2)}, P^{(3)}, \ldots\ldots \quad P^{(N-2)}]((N-1)t\hbar \to 1;\; x_1, x_2, \ldots x_{N-1})}$$

$$(K > 0) \tag{A.23}$$

$$H_{N,0} = 1 \tag{A.24}$$

**Appendix B**

This appendix is added because there is a similarity between another expansion method of $\Pi_{i=1}^{N} F(x_i,\alpha,\tau)$ in genus 1 and a higher genus expansion of $\Pi_{i=1}^{N} F_g(x_i,\alpha,\tau)$ .

In Chapter 1, we expanded $\Pi_{i=1}^{N} F(x_i,\alpha,\tau)$ by $P(\alpha)$, $P^{(1)}(\alpha)$, $P^{(2)}(\alpha)$, $\ldots P^{(N-2)}(\alpha)$ as

$$\Pi_{i=1}^{N} F(x_i,\alpha,\tau) = \sum_{M=0}^{N} H_{N,M}(x_i) P^{(N-2-M)}(\alpha) = H_{N,N}(x_i) + H_{N,N-2}(x_i)P(\alpha) + H_{N,N-3}(x_i)P^{(1)}(\alpha) + H_{N,N-4}(x_i)P^{(2)}(\alpha) + \cdots + H_{N,0}P^{(N-2)}(\alpha) \quad . \tag{1.24}$$

In genus 1, there is another way of expansion, using a set of functions $P(\alpha)$, $P^{(1)}(\alpha)$, $P^2(\alpha)$, $P^{(1)}(\alpha)\,P(\alpha)$, $P^3(\alpha)$, $P^{(1)}(\alpha)P^2(\alpha)$ , $P^4(\alpha)$ , $P^{(1)}(\alpha)P^3(\alpha)$, … . The orders of the poles of $\alpha$ are 2, 3, 4, 5, 6, 7,8 ,9 … respectively. The expansion has the form

$$\Pi_{i=1}^{N} F(x_i,\alpha,\tau) = \overleftrightarrow{H}_{N,N}(x_i) + X_{N,N-2}(x_i)P(\alpha) + X_{N,N-3}(x_i)P^{(1)}(\alpha) + X_{N,N-4}(x_i)P^2(\alpha) + + X_{N,N-5}P^{(1)}(\alpha)P\ \ (\alpha) + X_{N,N-6}P^3(\alpha) + \cdots + X_{N,0}P^{\frac{N}{2}}(\alpha) \tag{B.1}$$

If $N$ is odd, the last term on the right hand side of (B.1) should be replaced by $X_{N,1}P^{(1)}(\alpha)P^{\frac{N-3}{2}}(\alpha)$.

According to the elliptic theory of one variable, (1.24) and (B.1) give the equivalent results, although the final expressions of the decomposition formula are superficially different[9].

If one adopts (1.24), it is possible to write down a compact closed form of the decomposition formula as (1.11). This is the reason why we adopt (1.24) in Chapter 1. Instead, the spin structure dependent part becomes a slightly more complicated form, $P^{(2K-2)}(\omega_\delta)$, which becomes a polynomial of the Pe function.

On the other hand, if we use the expansion form (B.1), the spin structure dependent part of the results becomes simpler. Since $P^{(1)}(\omega_\delta) = 0$ and $P^M(\omega_\delta) = e_\delta^M$, there is no need to consider rewriting $P^{(2K-2)}$ as a polynomial of $P$ .

Instead, the forms of the coefficients $X_{N,M}(x_i)$ become rather complicated, because the expansion base functions on the right hand side of eq.(B.1) contain products of Pe functions such as $P^K(\alpha)$ and $P^{(1)}(\alpha)P^K(\alpha)$ . Given the expansion form of $P$ and

[9] In the first version of [2], the author adopted the expansion (1.24). After that, the author thought (wrongly) that there is a logical defect in the expansion of the form (1.24) and rewrote [2] using the expansion (B.1). This was totally by my misunderstanding. As for [2], please see version 1 only.

$P^{(1)}$ in (A.8), the process of obtaining the coefficients $X_{N,M}(x_i)$ as in the proof of the decomposition formula in Chapter 1 is not so simple[10].

In genus 1, for example for $N = 6$, after setting $\alpha = \omega_\delta$ ,

$$S_\delta(z_1, z_2)S_\delta(z_2, z_3)S_\delta(z_3, z_4)S_\delta(z_4, z_5)S_\delta(z_5, z_6)S_\delta(z_6, z_1) = \prod_{i=1}^{6} S_\delta(x_i)$$
$$=\widetilde{\widetilde{H}}_{6,6} + X_{6,4}P \quad (\omega_\delta) + X_{6,2}P^2(\omega_\delta) + X_{6,0}P^4(\omega_\delta)$$
$$=\widetilde{\widetilde{H}}_{6,6} + X_{6,4}e_\delta + X_{6,2}e_\delta^2 + X_{6,0}e_\delta^4 \tag{B.2}$$

where

$$\widetilde{\widetilde{H}}_{6,6} = V_6 - 6G_4V_2 - 15G_6$$
$$X_{6,4} = V_4 - 9G_4$$
$$X_{6,2} = V_2$$
$$X_{6,0} = 1 \tag{B.3}$$

For large $N$, the form of $X_{N,M}(x_i)$ will be a complicated form, while the spin structure dependent terms will have simple forms such as $e_\delta, e_\delta^2, e_\delta^4, \dots$ .

In higher genus, as we saw in chapter 2, we have to consider the dimension of the function space spanned by the expansion base functions. Roughly speaking, as for the "even order" of poles of expansion base functions in in higher genus, the set of the products of Pe functions, $P_{IJ}(\alpha)$ , $P_{IJ}(\alpha)P_{KL}(\alpha)$, $P_{IJ}(\alpha)\, P_{KL}(\alpha)P_{MN}(\alpha)\dots$. will span the largest function space. Or these direct products of Pe functions should be included in the desired base functions, instead of $P_{IJKL}(\alpha)$, $P_{IJKLMN}(\alpha), \dots$. We will also need to prepare other functions which have the "odd order" poles. In genus 1, if the base functions of the expansion, $P(\alpha)$, $P^2(\alpha)$, $P^3(\alpha)$, $P^4(\alpha)$ ... are prepared as functions with even order poles, only one more, $P^{(1)}(\alpha)$ , is multiplied to them and odd order poles are prepared. In genus one, this is sufficient to prepare all the base functions.

In higher genus, if the following formula is valid as a generalisation of (A.17), the same procedure may be validated:

For odd $K$ $(K \geq 5)$, $P_{I_1 I_2 \dots I_K}(\alpha)$ can be expressed as follows:

$$P_{I_1 I_2 \dots I_K}(\alpha) = P_{I_1 I_2 I_3}(\alpha) \times [\, polynomial\ of\ P_{JK}\,] \tag{B.4}$$

The author does not know whether this is validated or not. If it is true, then only $P_{I_1 I_2 I_3}(\alpha)$ may be necessary to produce odd-order pole functions which multiply to give $P_{IJ}(\alpha)$ , $P_{IJ}(\alpha)P_{KL}(\alpha)$, $P_{IJ}(\alpha)\, P_{KL}(\alpha)P_{MN}(\alpha)\dots$ .

If (B.4) is not true, one may have to prepare

---

[10] In the determinant formula, there is no such difficulty. If one replace $P,\ P^{(1)},\ P^{(2)},\ P^{(3)}, \dots P^{(N-2)}$ in the formula (A.22) and (A.23) with $P,\ P^{(1)},\ P^2,\ P^{(1)}P,\ P^3,\ P^{(1)}P^2,\ P^{(4)}\dots$ until $P^{\frac{N}{2}}$ if N is even, $P^{(1)}P^{\frac{N-3}{2}}$ if N is odd, then one can naturally get the determinant form of $\widetilde{\widetilde{H}}_{N,N}$ and $X_{N,N-M}$ .

$P_{I_1I_2I_3}(\alpha)\, P_{JK}(\alpha) P_{LM}(\alpha)$ and $P_{I_1I_2I_3I_4I_5}(\alpha)\, P_{JK}(\alpha)$ as base functions. These functions with "odd order " poles may affect the forms of the expansion coefficients and so this should be clarified in the future.

In a sense, for higher genus cases, we are forced to generalise the expansion of the form (B.1), instead of (1.22). By setting $\alpha = \Omega_\delta$, the base functions that have factors of the Pe function $P_{I_1I_2 \ldots I_M}$ for odd $M$ will be zero, so only the direct products of Pe functions $P_{IJ}(\Omega_\delta)$ , $P_{IJ}(\Omega_\delta)P_{KL}(\Omega_\delta)$, $P_{IJ}(\Omega_\delta)\, P_{KL}(\Omega_\delta)P_{MN}(\Omega_\delta)\ldots$ will remain non-zero in the final form of the decomposition formula of products of fermion correlation functions with cyclic conditions, as the most general result of spin structure dependent terms.

All the above is by analogy with the case of genus 1. In the case of higher genus, the situation can arise that higher rank tensors $P_{I_1I_2 \ldots I_K}$ for even integers of K>2 should be included in the base functions of the expansion and these are not decomposable as the product of $P_{AB}$ . In such cases the spin sum method described in chapter 2 can't be applied. On p.116 – p.129 of ref. [16] describes for the case of genus 2 and hyperelliptic genus 3 that the rank 4 tensors $P_{ABCD}$ can be written as polynomials of $P_{EF}$ . In these cases, by differentiating $P_{ABCD}$ arbitrary many times, $P_{I_1I_2 \ldots I_K}$ can be written for any K only by rank 2 and rank 3 tensors $P_{EF}$ and $P_{GHI}$ .
In other cases, however, it is not clear whether the higher rank tensors $P_{ABCD}$ can be written as polynomials of $P_{EF}$ . This is an open question for most of the cases and it would be important to pursue efficient spin structure sum methods in higher genus.

## Appendix C Characteristics of all spin structures in hyper-elliptic genus g case

For any genus $g$, if $e_{2g+2}$ is fixed at $\infty$, all characteristics will be classified as follows, in the hyperelliptic case [10][11].
Choose any combination of $g-k$ number of branch points out of $2g+1$ number of branch points excluding $e_{2g+2}$, for a fixed value of integer $k$. ( $k = 0,1,2, \ldots g$.)
As in (2.6), calculate the following integral for each of $g-k$ number of branch points :

$$\Omega_j = (2M)^{-1} \int_{\infty}^{e_j} \omega_I = a_j\ T + b_j \ , \tag{2.6}$$

and then add all these $g-k$ number of results, modulo 2 in the numerators of $a_j$ and $b_j$. We denote the result as $\Omega_{i_1 i_2 \ldots i_{g-k}}$ .

The number of characteristics of the type $\left[\, R + \Omega_{i_1 i_2 \ldots i_{g-k}} \right]$ is $\binom{2g+1}{g-k}$.

In the case of the odd characteristic, the meaning of "non-singular" is different from that in the case of even characteristic. Non-singular odd means that the first differential of the zero characteristic theta function at the origin is not zero for at least one of the variables.

The characteristics of all spin structures are given as follows: ( further explanation will be given below)

$\left[\ R+\ \Omega_{i_1 i_2 \dots i_g}\right]$ : non singular and even, order 0 .

$\left[\ R+\ \Omega_{i_1 i_2 \dots i_{g-1}}\right]$ *and* $\left[\ R+\ \Omega_{i_1 i_2 \dots i_{g-2}}\right]$ : non singular and odd, order 1.

$\left[\ R+\ \Omega_{i_1 i_2 \dots i_{g-3}}\right]$ *and* $\left[\ R+\ \Omega_{i_1 i_2 \dots i_{g-4}}\right]$ : singular and even, order 2.

$\left[\ R+\ \Omega_{i_1 i_2 \dots i_{g-5}}\right]$ *and* $\left[\ R+\ \Omega_{i_1 i_2 \dots i_{g-6}}\right]$ : singular and odd, order 3.

$\left[\ R+\ \Omega_{i_1 i_2 \dots i_{g-7}}\right]$ *and* $\left[\ R+\ \Omega_{i_1 i_2 \dots i_{g-8}}\right]$ : singular and even, order 4.,

……

$[\,R\,]$ : even (odd) if $\frac{g(g+1)}{2}$ is even (odd). singular except $g=1,2$. order $\left[\frac{g+1}{2}\right]$

(C.1)

The characteristic of Riemann constants of genus $g$, $[\,R\,]$, is given by a pair of the following vectors in hyperelliptic case :

$$\Delta_a = \left(\frac{g}{2},\ \frac{g-1}{2},\ \frac{g-2}{2}, \dots\ \frac{1}{2}\right),\quad \Delta_b = \left(\frac{1}{2},\ \frac{1}{2}, \dots\ \frac{1}{2}\right) \qquad (\mathit{mod}\ 2\quad \mathit{in\ the\ numerator})$$

(C.2)

In eq.(2.27) of ref.[11], partitions of $2g+2$ branch points are described. In that equation, $g+1-2m$ number of branch points are selected from $2g+2$ points, and the corresponding $\Omega_j$ are calculated, and then summed. The characteristics are even (odd) if $m$ is even (odd). The $m$ is equal to the order of zero of zero-characteristic theta function $\vartheta\left(\ R+\ \Omega_{i_1 i_2 \dots i_{g-k}}\ +v\right)$ at the origin $v=0$, by Riemann's vanishing theorem. (Here $k=0$ *if* $m=0$, *and* $k=2m-1$ *or* $2m$ *if* $m>0$ . )

The first line of (C.1) is the case $m=0$, the second line is the case $m=1$, the third line is the case $m=2$, …. .

The result of (C.1) can be easily understood if we think as follows.

If $e_{2g+2}$ is not fixed at $\infty$, for a fixed value of $m$, the spin structure corresponds to the way of choosing $g+1-2m$ branch points out of $2g+2$ branch points. When $e_{2g+2}$ is fixed at $\infty$, then the number of ways of choosing $g+1-2m$ branch points out of $2g+2$ branch points can be split into two ways as

$$\binom{2g+2}{g+1-2m} = \binom{2g+1}{g+1-2m} + \binom{2g+1}{g-2m} \ . \tag{C.3}$$

The first term on the right hand side of (C.3), $\binom{2g+1}{g+1-2m}$, corresponds to the case that the point $e_{2g+2}(=\infty)$ is not included in the choice of $g+1-2m$ number of points out of $2g+2$ points. The second term corresponds to the case where the point $e_{2g+2}(=\infty)$ is included in the choice of $g+1-2m$ number of points from $2g+2$ points.

For example, for $m=1$, $\binom{2g+1}{g+1-2m}$ corresponds to $\left[\, R+\ \Omega_{i_1 i_2 \dots i_{g-1}} \right]$, while $\binom{2g+1}{g-2m}$ corresponds to $\left[\, R+\ \Omega_{i_1 i_2 \dots i_{g-2}} \right]$.

For $m=2$, these two correspond to $\left[\, R+\ \Omega_{i_1 i_2 \dots i_{g-3}} \right]$, and $\left[\, R+\ \Omega_{i_1 i_2 \dots i_{g-4}} \right]$ respectively, and so on.

This shows the reason why in (C.1), where $e_{2g+2}$ is fixed at $\infty$, the words odd and even are swapped every two types of characteristics represented by $[\, R+\ \Omega\ \ ]$, except the non-singular and even case.

If we count the sum of the ways of selecting $k$ number of branch points out of $2g+1$ points for all values of $k$, we see that it reproduces a well-known result on the total number of spin structure characteristics $2^{2g}$. By using a formula

$$\binom{n+1}{r} = \binom{n}{r} + \binom{n}{r-1} \ (for\ 1 \le r \le n)\ , \tag{C.4}$$

$$\sum_{k=0}^{g}\binom{2g+1}{k} = 1+\sum_{k=1}^{g}\binom{2g+1}{k} = 1+\sum_{k=1}^{g}\binom{2g}{k} + \sum_{k=1}^{g}\binom{2g}{k-1}$$

$$= 1+\sum_{k=1}^{g}\binom{2g}{k} + \sum_{k=g+1}^{2g}\binom{2g}{k} = \sum_{k=0}^{2g}\binom{2g}{k} = (1+1)^{2g} = 2^{2g} \ . \tag{C.5}$$

Eq.(C.1) reproduces all exemplary cases of genus 2, 3, 4 in section 2.1 or in [13].

## Appendix D Miscellaneous issues

### D-1 Weierstrass relations

Weierstrass relations at genus one (1.46) can be proved as follows.

In the definition of Pe function:

$$P(x) = -2\eta_1 - \frac{\partial^2}{\partial x^2} \ln \theta_1(x,\tau) = -2\eta_1 - \left[ \frac{\theta_1^{(2)}(x)}{\theta_1(x)} - \left( \frac{\theta_1^{(1)}(x)}{\theta_1(x)} \right)^2 \right] \quad , \tag{D.1}$$

replace the function $\frac{\theta_1^{(k)}(\omega_\delta)}{\theta_1(\omega_\delta)}$ with $\theta_{\delta+1}^{(k)}(0)$ for $k = 1,2$ , by using (1.20) :

$$\frac{\theta_1(x+\omega_\delta)}{\theta_1(\omega_\delta)} = exp(-\pi i\, a_\delta\, x)\, \frac{\theta_{\delta+1}(x)}{\theta_{\delta+1}(0)} \ . \tag{1.20}$$

Then, because $\theta_{\delta+1}(x)$ are even, we have the Weierstrass relations for $\delta = 1,2,3$.

As for the higher -genus Weierstrass relations (2.26) in hyper-elliptic case, though it seems that the proof was not explicitly described in the original paper [12] nor anywhere else, it may be derived in a similar way as in the case of genus one.

Let $\theta_R^{(k)}(\Omega_\delta)$ be the $k$-number of differentiations with respect to any variables evaluated at $u = \Omega_\delta$ and $\delta$ represents non-singular and even spin structures.

In the definition of higher-genus Pe functions, $\frac{\theta_R^{(1)}(\Omega_\delta)}{\theta_R(\Omega_\delta)}$, $\frac{\theta_R^{(2)}(\Omega_\delta)}{\theta_R(\Omega_\delta)}$ can be replaced with $\theta\ [\delta](0)$, $\theta^{(1)}[\delta](0), \theta^{(2)}[\delta](0)$ and constants which relate to the differentiations on the factor $\exp\left(-2\pi i \left(\sum_{k=1}^{g} a_{i_k}\ \right)\cdot u\right)$ , by eq.(2.18) :

$$\frac{\theta_R(u+\Omega_\delta)}{\theta_R(\Omega_\delta)} = \exp\left(-2\pi i \left(\sum_{k=1}^{g} a_{i_k}\ \right)\cdot u\right)\ \frac{\theta[\delta](u)}{\theta[\delta](0)} . \tag{2.18}$$

Using the fact that $\theta[\delta](u)$ is even, since $\theta^{(1)}[\delta](0) = 0$, we have

$$\frac{\partial^2}{\partial u_I\, \partial u_J} \ln \theta_R\,(u)\bigg|_{u=\Omega_\delta} = \frac{\partial_I \partial_J\, \theta[\delta](0)}{\theta[\delta](0)} \tag{D.2}$$

By the definition of Pe function :

$$P_{JK} = -\frac{\partial^2}{\partial u_J \partial u_K} \ln\theta_R((2M)^{-1}u) - 2\eta(2M)^{-1}{}_{JK} \ ,$$

the differentiations with respect to the components of $u = (u_1, u_2, .. u_I, .. u_g\ )$ will have a little complicated form, but since $\frac{\partial_I\, \theta[\delta](0)}{\theta[\delta](0)} = 0$ for any $I$ , the higher genus Weierstrass relations (2.26) follow.

In the proof of $P_{IJK}(\Omega_\delta) = 0$ , the same logic can be applied for genus $g$ , by differentiating both sides of (2.18) one more time.

**D-2 Arbitrary number of differentiations of the Pe function valued at the non-singular and even half periods**

**D-2-1 Genus one**

As is well known, for genus one,

$$P^{(1)}(\omega_\delta) = 0 = P^{(2k-1)}(\omega_\delta)$$

for any positive integer $k$. This can be verified by (A.16) and (A.17).

This can also be confirmed step by step by differentiating the both sides of (1.20) and substitute those into $P^{(k)}(\omega_\delta)$ for $k = 1.2.3, \dots$ .

By (1.20) above, each of the values of $\frac{\theta_1^{(k)}(\omega_\delta)}{\theta_1(\omega_\delta)}$ for $k = 1.2.3, \dots$ depends on the value of $a_\delta$ , but the expressions of $P^{(2k)}(\omega_\delta)$ don't depend on the explicit values of $a_\delta$ for any $k$. The $P^{(2k)}(\omega_\delta)$ can be expressed as a polynomial of $\frac{\theta_{\delta+1}^{(m)}(0)}{\theta_{\delta+1}(0)}$ for $m = 2,4, \dots . k$ , and the form of the polynomial is the same for any $\delta = 1,2,3$ , if the half periods are set as $\omega_1 = \frac{1}{2}$, $\omega_2 = -\frac{1+\tau}{2}$, $\omega_3 = \frac{\tau}{2}$ .

**D-2-2 Higer genus**

We write (2.18) as

$$\frac{\theta_R(u+\Omega_\delta)}{\theta_R(\Omega_\delta)} = \exp(\mathrm{L}\cdot\mathrm{u})\ h(u) \quad , \quad \text{where} \qquad h(u) = \frac{\theta[\delta](u)}{\theta[\delta](0)} . \tag{D.3}$$

The $u$ is a set of $g$ number of variables, $u_1, u_2, \dots, u_g$ .

When $L$ has the form $L = \cdots + Au_i + \cdots$ where $A$ is a numerical factor and $u_i$ is one of the components of $u$, we call $u_i$ " the variable corresponding to the coefficient $A$. "

Define $h^A(u)$ as the derivative of $h$ with respect to the variable corresponding to $A$ .

We also define any number of differentiations of $h$ as $h^{AB}(u)$, $h^{ABC}(u)$, $h^{ABCD}(u)$, .... .

The variables corresponding to $A, B, C, D$ .... allow duplicated choice.

If we differentiate (D.3) three times for example, we have

$$\frac{\theta_R^{A_1A_2A_3}(u+\Omega_\delta)}{\theta_R(\Omega_\delta)} = A_1A_2A_3e^{\mathrm{L\cdot u}}\ h(u) + A_1A_2e^{\mathrm{L\cdot u}}\ h^{A_3}(u) + A_2A_3e^{\mathrm{L\cdot u}}\ h^{A_1}(u) + A_3A_1e^{\mathrm{L\cdot u}}\ h^{A_2}(u)$$
$$+ A_1e^{\mathrm{L\cdot u}}\ h^{A_2A_3}(u) + A_2e^{\mathrm{L\cdot u}}\ h^{A_3A_1}(u) + A_3e^{\mathrm{L\cdot u}}\ h^{A_1A_2}(u) + e^{\mathrm{L\cdot u}}\ h^{A_1A_2A_3}(u) \tag{D.4}$$

Setting $u = 0$, since $h^{A_1}(0) = h^{A_1A_2A_3}(0) = 0$ because $\theta[\delta](u)$ is even, we have

$$\frac{\theta_R^{A_1A_2A_3}(\Omega_\delta)}{\theta_R(\Omega_\delta)} = A_1A_2A_3 + A_1h^{A_2A_3}(0) + A_2h^{A_3A_1}(0) + A_3h^{A_1A_2}(0) \tag{D.5}$$

By induction, we have, since $h^{(ODD)}(0) = 0$ due to the fact that $h$ is even,

$$\frac{\theta_R^{A_1A_2A_3\cdots A_M}(\Omega_\delta)}{\theta_R(\Omega_\delta)} = A_1A_2\cdots A_M + \{ A_3\cdot A_4\cdots A_M\cdot h^{A_1A_2}(0) + \ combinations \}$$

$$+\{ A_5\cdot A_6\cdots A_M\cdot h^{A_1A_2A_3A_4}(0) + \ combinations \ \} + \ \cdots \tag{D.6}$$

On the other hand, define

$$f^A(u) \stackrel{\text{def}}{=} \frac{\theta_R^{A}(u)}{\theta_R(u)} = \frac{d}{du_i}\ ln\theta_R(u) = \ [ln\theta_R(u)]^A \tag{D.7}$$

Differentiating more of the both sides of $f^A(u)\theta_R(u) = \theta_R^{A}\ (u)$ and divided by $\theta_R(u)$ , and setting $u = \Omega_\delta$ , we have by (D.6) , for example,

$$f^{A_1}(\Omega_\delta) = \ A_1 \ , \tag{D.8}$$

$$f^{A_1A_2}(\Omega_\delta) = \ h^{A_1A_2}(0)\,, \tag{D.9}$$

$$f^{A_1A_2A_3}(\Omega_\delta) = \ 0\ , \tag{D.10}$$

$$f^{A_1A_2A_3A_4}(u) + \ f^{A_1A_2A_3}(u)\ \frac{\theta_R^{A_4}(u)}{\theta_R(u)} + f^{A_1A_2A_4}(u)\ \frac{\theta_R^{A_3}(u)}{\theta_R(u)} + f^{A_1A_3A_4}(u)\frac{\theta_R^{A_2}(u)}{\theta_R(u)}$$

$$+ \ f^{A_1A_2}(u)\ \frac{\theta_R^{A_3A_4}(u)}{\theta_R(u)} + f^{A_1A_3}(u)\ \frac{\theta_R^{A_2A_4}(u)}{\theta_R(u)} + f^{A_1A_4}(u)\ \frac{\theta_R^{A_2A_3}(u)}{\theta_R(u)} + f^{A_1}(u)\frac{\theta_R^{A_2A_3A_4}(u)}{\theta_R(u)}$$

$$= \frac{\theta_R^{A_1A_2A_3A_4}(u)}{\theta_R(u)} \tag{D.11}$$

When setting $u = \Omega_\delta$ in (D.11), all terms which depend on explicit values of $A_1, A_2, A_3, A_4$ such as $A_1A_2h^{A_3A_4}(0)$ etc. disappear after using (D.6), and we have,

$$f^{A_1A_2A_3A_4}(\Omega_\delta) = h^{A_1A_2A_3A_4}(0) - \big( h^{A_1A_2}(0)h^{A_3A_4}(0) + h^{A_1A_3}(0)h^{A_2A_4}(0) + \\ + h^{A_1A_4}(0)h^{A_2A_3}(0)\big). \tag{D.12}$$

By induction, for $k = 1,2,3,\ldots$ , we have

$$f^{A_1}(\Omega_\delta) = \ [ln\theta_R(\Omega_\delta)]^{A_1} = A_1 \tag{D.13}$$

$$f^{A_1A_2\cdots A_{2k+1}}(\Omega_\delta) \ = \ [ln\theta_R(\Omega_\delta)]^{A_1A_2\cdots A_{2k+1}} \ = \ 0 \tag{D.14}$$

$$f^{A_1A_2\cdots A_{2k}}(\Omega_\delta) = \ \ [ln\theta_R(\Omega_\delta)]^{A_1A_2\cdots A_{2k}} = h^{A_1A_2\cdots A_{2k}}(0) - Y \tag{D.15}$$

where

$$Y = [\ h^{A_1A_2}(0)h^{A_3A_4\cdots A_{2k}}(0) + combinations] \ + [h^{A_1A_2A_3A_4}(0)h^{A_5A_6\cdots A_{2k}}(0) + combinations] \\ + [h^{A_1A_2A_3A_4A_5A_6}(0)h^{A_7A_8\cdots A_{2k}}(0) + combinations] \ldots. + [\ h^{A_1A_2\cdots A_k}(0)h^{A_{k+1}A_{k+2}\cdots A_{2k}}(0) + \\ combinations] \ \ . \tag{D.16}$$

The form of $f^{A_1A_2\cdots A_{2k}}(\Omega_\delta)$ , (D.15) shows an explicit way of how $[ln\theta_R(\Omega_\delta)]^{A_1A_2\cdots A_{2k}}$ can be written as a polynomial of $\frac{\theta^{(EVEN)}[\delta](\Omega_\delta)}{\theta[\delta](0)}$ . The result does not depend on any of the explicit numerical values of $A_1, A_2, \ldots A_{2k}$ .

In the definition of Pe function, $ln\theta_R$ is the function of linear combinations of the components of $u$ , $(2M)^{-1}u$ :

$$P_{JK} = -\frac{\partial^2}{\partial u_J \partial u_K} ln\theta_R((2M)^{-1}u) - 2\eta(2M)^{-1}{}_{JK} \ , \qquad P_{I_1 I_2 \dots I_N} = \frac{\partial^{N-2}}{\partial u_{I_3} \partial u_{I_4} \dots \partial u_{I_{N1}}} P_{I_1 I_2} \ .$$

When we differentiate $P_{JK}$ with respect to the components of $u$ and then set $u$ equal to $\Omega_\delta$, the result will be mixture combinations of $f^{A_1 A_2 \cdots}(\Omega_\delta)$ . But this does not affect the fact that $[ln\theta_R((2M)^{-1}u)]^{A_1 A_2 \dots A_{2k+1}}$ becomes zero at $u = \Omega_\delta$ because the result is a linear combination of $f^{A_1 A_2 \cdots}(\Omega_\delta)$ with the same total number of the differentiations and (D.14) does not depend on the choice of corresponding variables.
Therefore we have

$$P_{I_1 I_2 \dots I_{2k-1}}(\Omega_\delta) = 0 \quad \text{if } k \text{ is any positive integer.}$$

**Acknowledgements**

First of all, the author is sincerely grateful to the referee for correctly pointing out a number of errors in the earlier versions of the manuscript and pointing the way to the solutions in several respects.

The author also thanks to Prof. A.Morozov for his kind help in the process of submitting ref.[2] to arXiv.

This paper is dedicated to Ms. Shala-Ni, who died by a car accident on Feb. 21 2021 at her age fifteen.

A.G. Tsuchiya e-mail colasumi_0506@yahoo.co.jp